\documentclass{aa}  
\usepackage{graphicx,amsmath}
\usepackage{txfonts}
\begin{document}
\newcommand{\mearth}{M_\oplus}
   \title{On the evolution of multiple low mass planets embedded in a circumbinary 
disc}


   \author{A. Pierens
          \and
          R.P Nelson
          }

   \offprints{A. Pierens}

   \institute{Astronomy Unit, Queen Mary, University of London, Mile End Rd, London, E1 4NS, UK\\
     \email{a.pierens@qmul.ac.uk}       
             }

   \date{Received September 15, 1996; accepted March 16, 1997}

 
  \abstract
{Previous work has shown that the tidal interaction between a binary system
and a circumbinary disc leads to the formation of a large inner cavity
in the disc.
Subsequent formation and inward migration of a low mass planet causes
it to become trapped at the cavity edge, where it orbits
until further mass growth or disc dispersal. The question of how
systems of multiple planets in circumbinary discs evolve has not
yet been addressed.}
   {
We present the results of hydrodynamic simulations of multiple low mass 
planets embedded in a circumbinary disc. The aim is to examine their long
term evolution as they approach and become trapped at
the edge of the tidally truncated inner cavity.
}
  {A grid--based hydrodynamics code was used to compute simulations of
   2D circumbinary disc models with embedded planets. The 3D evolution of
   the planet orbits was computed, and inclination damping due to the
   disc was calculated using prescribed forces. We present a suite of
   simulations which study the evolution of pairs of planets migrating
   in the disc. We also present the results of hydrodynamic simulations 
   of five-planet systems, and study their long term evolution after disc 
   dispersal using a N-body code.}
   {
   For the two-planet simulations
   we assume that the innermost planet has migrated to the edge of the
   inner cavity and remains trapped there, and study the subsequent
   evolution of the system as the outermost planet migrates inward.
   We find that the outcomes largely depend on the mass ratio $q=m_i/m_o$, 
   where $m_i$ ($m_o$) is the mass of the innermost (outermost) planet.
   For $q<1$, planets usually undergo dynamical scattering or orbital
   exchange. For values of $q>1$ the systems reach
   equilibrium configurations in which
   the  planets are locked into mean motion resonances, and remain
   trapped at the edge of the inner cavity without further migration.
   Most simulations of five-planet systems we performed resulted
   in collisions and scattering events, such that only a single
   planet remained in orbit about the binary.
   In one case however, a multiplanet resonant system was
   found to be dynamically stable over long time scales, 
   suggesting that such systems may be observed in planet searches
   focussed on close binary systems.}
   {}


   \maketitle
%

\section{Introduction}

At the time of writing approximately 250 extrasolar planets have been
discovered, of which about 30 are members of binary or multiple stellar
systems (Eggenberger et al. 2004; Mugrauer et al. 2007).
Most of the planets found in binaries orbit around one of the 
stellar component in so--called S--type orbits, and the majority
of binaries harbouring planets have orbital separations $a_b \ge 100$ AU.
There are, however, exceptions to these cases. The short period
binary systems Gliese 86,
$\gamma$ Cephei and HD41004 have $a_b \sim 20$ AU, and contain planets
orbiting at 1 - 2 AU from the central star (Eggenberger et al. 2004; 
Mugrauer \& Neuhauser 2005). Although there are no known planets which
orbit around both stellar companions in a binary system consisting of
main sequence stars (i.e. circumbinary planets on so--called P-type orbits),
there are two systems which indicate that the formation of circumbinary
planets is feasible. The first is the circumbinary planet with mass
$m_p=2.5$ M$_{J}$ and orbital radius 23 AU observed in the radio pulsar
PSR 1620-26 (Sigurdsson et al. 2003). The second is the $m_p=2.44$ M$_{J}$ planet orbiting about
the system composed of the star HD202206 and its 17.4 M$_J$ brown dwarf 
companion (Udry et al. 2002). The lack of observed circumbinary planets is
probably due to the fact that short period binaries are usually rejected
from observational surveys.

The observation of planets in binary systems is consistent with detections
of circumstellar discs in binary systems. Several circumbinary discs have 
been detected around spectroscopic binaries such as DQ Tau, AK Sco, and GW Ori.
The circumbinary disc around GG Tau has been directly imaged 
(Dutrey et al. 1994), revealing the presence of a tidally truncated inner
cavity generated by the central binary. The existence of these circumbinary
discs opens up the possibility of circumbinary planets forming.
Combining the observations of circumbinary discs, with the fact
that $\sim 50 \%$ of solar-type stars are members of binaries (Duquennoy \&
Mayor 1991), suggests that circumbinary planets are probably common.

To date there has been a relatively modest amount of theoretical
work examining planet formation in binary systems. Results from
  previous studies 
suggest that planetesimal accretion should be possible
in regions of both circumstellar (Marzari \& Scholl 2000; Thebault et
al. 2006) and circumbinary discs (Moriwaki \& Nakagawa 2004; Scholl et
al. 2007). Quintana \& Lissauer (2006) simulated the late stages of
terrestial planet formation in circumbinary
discs. They found that planetary systems similar to those around
single stars can be formed around binaries, provided the ratio
of the binary apocentre distance to planetary orbit is
$\le 0.2$. In general binaries with larger maximum separations lead to
planetary systems with fewer planets.

The evolution of a low mass planetary core embedded in a
circumbinary disc was investigated recently by Pierens \& Nelson (2007) 
(hereafter referred to as Paper I). This work examined the migration and 
long term orbital evolution of planets with masses of 
$m_p=$ 5, 10 and 20 $M_\oplus$  under the action of disc torques. It was found
that the inward drift of a planet undergoing type I migration is  
stopped at the edge of the cavity formed by the binary.
This halting of migration is due to positive corotation torques operating
which can counterbalance negative Lindblad
torques. Such an
effect is known to be at work in accretion disc regions where there is a 
strong positive gradient of the surface density (Masset et al. 2006). 
Interestingly, Pierens \& Nelson (2007) showed that the stopping of 
migration in circumbinary discs occurs in a region of long-term dynamical
stability, suggesting that such planets may be able to survive
there over long times, or at least remain in the disc for long enough
to form a gas giant planet. The evolution of giant planets in circumbinary
discs was considered by Nelson (2003).

In this paper, we extend the model presented in Paper I by considering the
evolution of multiple planets embedded in a circumbinary disc. 
Here, we wish to examine how multiple planets interact with each other 
if they form at large distance from the binary and successively migrate 
toward the cavity edge. In particular, we want to look at whether or not 
the trapping of a planet at the cavity edge, and the subsequent
migration of additional planets to its vicinity lead to growth
of the planet through collisions, the formation of mean motion resonances,
or destablisation of the system through gravitational scattering. 

To address
these issues, we first consider a system which consists of a pair of
planets with masses of $m_p=$ 5, 10 and 20 $M_\oplus$. 
We assume that one planet is trapped at the edge of the cavity while 
the outermost planet migrates in from  larger radius. 
The results of the simulations show that the final outcome of such a system
generally depends on the mass ratio $q=m_i/m_o$ (where $m_i$ is the mass
of the inner planet and $m_o$ is the mass of the outer planet). 
Interestingly, we find that
systems with $q\ge 1$ can reach a steady state such that the planets
are locked into resonance and remain trapped at the cavity edge. 
Most of the systems with $q< 1$, however,  are unstable and lead to events such
as scattering or dynamical exchange. 

We performed a second set of simulations consisting of
five-planet systems embedded in a circumbinary disc. 
Of three simulations performed, two resulted in a single planet
orbiting around the binary because of collisions and scattering events.
The remaining simulation resulted in a three-planet system remaining,
with all planets in mutual mean motion resonances. This configuration
was found to be stable over long time scales.

This paper is organized as follows. In section 2 we describe the
physical model and the numerical method. In section 3 we describe the
results of simulations aimed at studying the evolution of pairs of planets embedded in a circumbinary disc. We then present in section 4 the simulations of five-planet systems embedded in a circumbinary disc. We finally discuss our results
 and present our conclusions in section 4.


\section{Physical model and numerical setup}
\subsection{Disc and planet evolution}

As in Paper I, we adopt a 2D disc model for which we assume no vertical
motion. The equations governing the disc evolution are
described in detail in Paper I and therefore will not be discussed
here. \\
In Paper I, the planet orbit and the disc midplane were
assumed to be coplanar. However, the simulations presented here
examine the evolution of multiple planets which can strongly
interact with each other as their orbits converge, leading
 eventually to close encounters. During such an event, a planet may receive a significant component of acceleration in the
 vertical direction, reducing thereby the interaction with both the
 disc and other planets.  As a consequence, we decided here to
 use a model in which planets can evolve in the $z$ direction as well.  With respect to coplanar orbits, this will also reduce
the collision rate between planets, increasing thereby the time
during which planets can strongly interact.\\
In the work presented here, each planet can interact with every
other one and with the disc. The latter interaction is expected to
lead not only to orbital migration but also to eccentricity and
inclination damping. The gravitational potential of the disc is
calculating using the following expression:

\begin{equation}
{\Phi}_{d}=-G \int_S\frac{\Sigma({\bf r}') d{\bf r}'}{\sqrt{r'^2+r_p^2-
2r'r_p\cos(\phi'-\phi_p)+z_p^2+\epsilon^2}}
\label{eq:pot}
\end{equation}
where $\Sigma$ is the disc surface density and where the integral is
performed over the disc surface. $r_p$, $\phi_p$ and $z_p$ are
respectively the radial, azimuthal and vertical coordinates of the
$p^{th}$ planet. $\epsilon$ is a smoothing parameter  which is
set to $\epsilon=0.6H$, where $H$ is the local disc scale
height. Under the action of this gravitational potential, each
planet undergoes both orbital migration and eccentricity
damping. However, because of the 2D disc model we use here, bending waves
cannot be launched in the disc, and so there is no disc induced damping of
inclination.
To model the latter we follow Tanaka \& Ward (2004)
and mimic the effect of bending waves by applying to each
planet a vertical force $F_z$ given by:
\begin{equation}
F_z=\beta\frac{m_p\Sigma_p\Omega}{c_s^4}(2A_z^cv_z+A_z^sz\Omega),
\end{equation}
where $c_s$ is the sound speed and where $\Omega$ and $\Sigma_p$ are
respectively  the Keplerian angular velocity and the disc surface
density at the position of the planet. $A_z^c$ and $A_z^s$ are
dimensionless coefficients which are set to  $A_z^c=-1.088$ and
$A_z^s=-0.871$ (Tanaka \& Ward 2004), and $\beta$ is a free
parameter which is chosen such that the inclination damping $t_i $ timescale
obtained in the simulations is approximatively equal to the eccentricity
damping timescale $t_e$. Test simulations show that choosing
$\beta=0.33$ give similar values for $t_i$ and $t_e$. We adopt
therefore this value for this work.  Here, it is worthwhile to notice that according
to the linear theory (Tanaka \& Ward 2004), a small value of the
planet inclination $(i_p\ll H/R)$ is not expected to affect the
migration rate of the planet. For large values of $i_p$ however,
migration rates may be moderately slowed down (Cresswell et al. 2007)
because the interaction with the disc is reduced. Such an effect is accounted for in an approximate manner by Eq. \ref{eq:pot}.

\subsection{Numerical setup}
\subsubsection{Numerical method}
The simulations presented here were performed using the hydrodynamic
code GENESIS. This code employs a second-order-accurate method that computes
advection using  the monotonic transport algorithm (Van Leer
1977). Details about GENESIS are given in Paper I.
All the runs use $N_r=256$ radial grid cells uniformly distributed
between $r_{in}=0.5$ and $r_{out}=6$ and $N_\phi=380$ azimuthal grid
cells.\\
The evolution of the planets plus binary system  is performed using a
5th-order Runge-Kutta scheme (Press et al. 1992). In spite of the accuracy
of this integrator, experiments have shown that to ensure
energy conservation during close encounters, the time step size
$\Delta t$ used to make the system evolve should be smaller than the hydrodynamical time step $\Delta t_h$ based on
the CFL criterion (Stone \& Norman 1992). Following Cresswell \& Nelson (2006),
we ensure good energy conservation and accuracy by setting the 
time step size to
$\Delta t$=min$(\Delta t_h, \Delta t_{n1}, \Delta t_{n2}) $, where:
\begin{equation}
\Delta t_{n1}=\frac{2\pi}{400}\underset{p,p'}{min}
\left(\sqrt{\frac{|{ {\bf r}_{pp'}}|^3}{G(m_p+m_p')}}\right),
\end{equation}
and
\begin{equation}
\Delta t_{n1}=\frac{2\pi}{400}\underset{p,s}{min}
\left(\sqrt{\frac{|{ {\bf r}_{ps}}|^3}{G(m_p+m_s)}}\right).
\end{equation}
In the previous expressions,  ${\bf r}_{pp'}$ is the distance between
 the planets $p$ and $p'$ and  ${\bf r}_{ps}$ is the distance between
the planet $p$ and star $s$.  We note that throughout our
 simulations, the time step size is $\sim 1/800$ the binary orbital period.\\
As in Paper I, we adopt computational units in which the total mass of the binary is $M_*=1$, the gravitational constant is $G=1$, and the radius $r=2$ in the computational domain corresponds to 5 AU. The unit of time is $\Omega^{-1}=\sqrt{GM_*/a_b^3}$, where $a_b=0.4$ is the initial semimajor axis of the binary. This corresponds to an initial separation between the two stars of $\sim 1\;\text{AU}$.\\
In the simulations presented here, close encounters between two planets can result in a physical collision. Here, this is supposed to occur whenever the mutual distance $d_{pp'}$ between planets $p$ and $p'$ is less than $(3m_p/4\pi\rho)^{1/3}+(3m_p'/4\pi\rho)^{1/3}$, where $\rho$ is the mass density which we assume to be the same for each planet and equal to $\rho=3\; g.cm^{-3}$. If a collision is found to occur between the planets $p$ and $p'$, these are assumed to merge and are subsequently  substituted by a single body with mass $m_p+m_p'$. The position and velocity of the latter are set to the position 
and velocity of the centre of mass of the planets $p$ and $p'$. 
\subsubsection{Initial conditions}

As in Paper I, the disc aspect ratio $H/R$ is assumed to be constant and 
equal to $H/R=0.05$. We use also the ``alpha'' prescription of 
Shakura \& Sunyaev (1973) to model the disc anomalous kinematic viscosity 
$\nu=\alpha c_s H$, where $c_s$ is the isothermal sound speed and 
where $\alpha=10^{-4}$. The reason for choosing such a low 
$\alpha$ value is discussed in detail in Paper I, but is essentially because
a larger value causes rapid evolution of the binary orbit that would
prohibit the long simulations we present here.\\
In Paper I, we showed that both the disc and binary evolve toward a
near-steady state as they interact with each other. From the time this
equilibrium configuration is reached, the apsidal lines of the disc and binary
are aligned. Also, the disc structure and the orbital elements
of the binary remain essentially constant.  For example, we find that the
eccentricity of a binary with mass ratio $q_b=0.1$ and initial 
separation $a_b=0.4$
saturates at a value of $e_b\sim 0.08$. The simulations presented in
paper I of one planet interacting with a circumbinary disc were 
performed using this quasi-equilibrium state as initial conditions for 
the disc and binary. Depending on the model we consider, we adopt here a
similar approach when setting up our initial conditions:\\
i) In simulations of pairs of planets embedded in
a circumbinary disc, we restart the runs presented in Paper I  once the
planet is trapped at the cavity edge but with a second planet
evolving on a circular orbit with
 $a_p=2.5$ and $i_p=0.5^\circ$. The latter is allowed to 
interact with the disc whose mass is $M_d\sim 0.01\; M_\star$ and with the other planet and binary. \\
ii) In simulations that evolve five-planet systems in a circumbinary disc, we
embed the planets in the disc once the latter and the binary have
reached a stationary state, as described in Paper I.
We set the innermost planet
at $a_p=1.8$ and then calculate the initial location of the others by
asssuming that two adjacent bodies $p$ and $p'$ are separated by $\sim 5
\;R_{mH}$, where $R_{mH}$ is the mutual Hill radius defined by:
\begin{equation}
R_{mH}=\left(\frac{m_p+m_{p'}}{3M_\star}\right)^{1/3}\left(\frac{a_p+a_{p'}}{2}\right).
\end{equation}
Each  body is assumed to initially evolve on a circular orbit
with $i_p=0.5^\circ$. Note
that the initial separation between planets we adopt is greater that the
critical value of $\sim 3.46\; R_{mH}$ below which rapid instability
occurs for two planets on initially circular orbits (Gladman 1993). 

\section{Evolution of pairs of planets embedded in circumbinary
  discs}
In paper I we considered planets with masses $m_p$= 5, 10 and
20 $M_\oplus$. Therefore, we adopt here the same values for the mass
$m_i$ of the innermost planet which is assumed to be trapped at the
cavity edge of the disc. For each value of $m_i$, we have performed two or three runs
for which $m_o$ was varied between $5 \le m_o \le 20
\;M_\oplus$.  Table \ref{table1} gives the values of $m_i$, $m_o$ and
 $q=m_i/m_o$  for each run. An
interesting feature of the results of these simulations is that
varying the value of $q$ can lead to different outcomes. Below,
we discuss in detail the different modes of evolution obtained in the simulations, and
how they change depending on whether  or not $q \ge 1$.
\begin{table}
\caption{The first column gives the run label, the second column gives
  the mass $m_i$ of the inner planet and the third column gives the
  mass $m_o$ of the outer planet. The fourth column gives the ratio $q=m_i/m_o$.}             
\label{table1}      
\centering          
\begin{tabular}{c c c c}     
\hline\hline       
Run & $m_i\;(\mearth)$ & $m_o\;(\mearth)$ & q\\ 
\hline                    
 Run1 & 10 & 10 & 1\\  
 Run2 & 10 & 5 & 2\\ 
 Run3 & 20 & 10 & 2\\ 
 Run4 & 20 & 5 & 4\\ 
 Run5 & 5 & 10 & 0.5\\ 
 Run6 & 10 & 20 & 0.5\\ 
 Run7 & 5 & 20 & 0.25\\ 
\hline                  
\end{tabular}
\end{table}
\subsection{Models with $q=1$}
\begin{figure*}
   \centering
  \resizebox{\hsize}{!} {\includegraphics{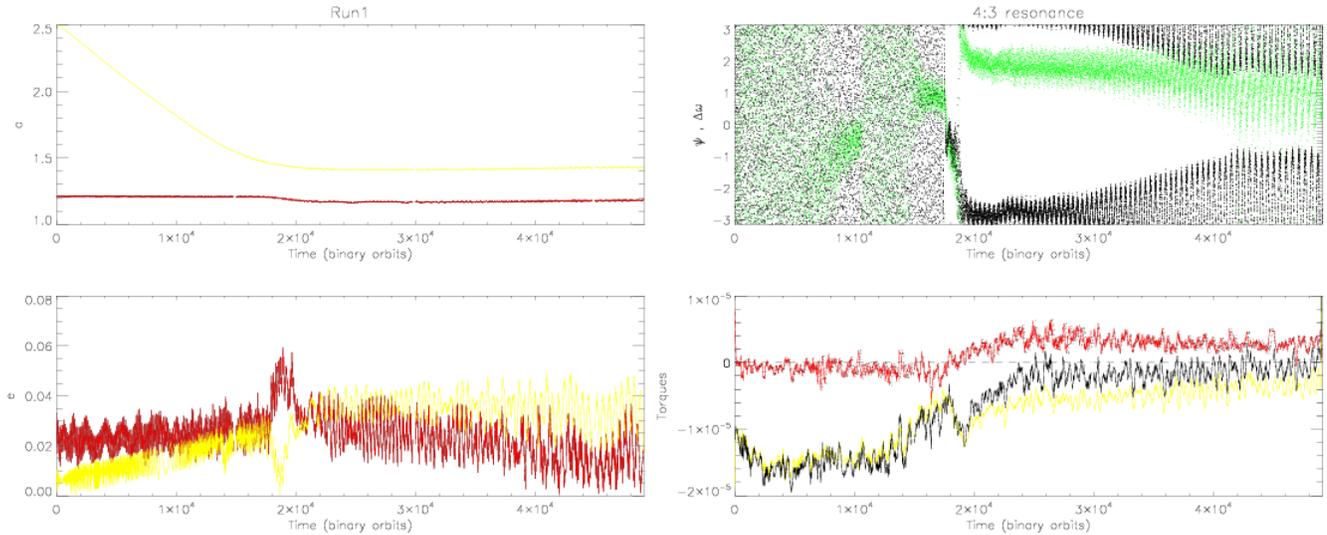}}
    \caption{{\it Left:} evolution of the semimajor axes (top
              panel) and eccentricities (bottom panel) of  planets with masses of $m_p=$ 10 $\mearth$. {\it Right:}
              the top panel shows the evolution of both the resonant angle
              $\psi=4\lambda_o-3\lambda_i-3\omega_i$ associated with
              the 4:3 resonance (black) and the apsidal angle
              $\Delta\omega=\omega_i-\omega_o$ (green). The bottom panel shows the evolution of the torques exerted by the disc on the
              innermost planet (red line) and on the outermost one (yellow line). The sum (red+yellow) is displayed with the black line and corresponds to the torques exerted on the whole system.
              }  
         \label{run1}
   \end{figure*}

The orbital evolution of a pair of  planets with masses of $m_p=$ 10
$M_\oplus$ is shown in the left panel of Figure \ref{run1}. At the
beginning of the simulation,  corotation torques cause the inner planet to remain trapped at the edge of the disc
cavity (see Paper I) located at  $r \sim 1.2$. The outer planet, which initially evolves on a
circular orbit with $a_o=2.5$, undergoes the usual type I migration
and drifts inward. As it migrates, its eccentricity $e_o$ slowly
increases due to the influence of the binary (see bottom left
panel of Figure \ref{run1}). Also, the distance between the two
bodies becomes smaller and smaller which can potentially leads to the
formation of  mean motion resonances (Papaloizou \&
Szuszkiewicz 2005). Here, we find that the two planets become captured into the 4:3
resonance at $t\sim 1.9\times 10^4$ binary orbits. The top right panel
of Figure \ref{run1} displays the evolution of both the apsidal angle
$\Delta\omega=\omega_i-\omega_o$ and the resonant angle
$\psi=4\lambda_o-3\lambda_i-3\omega_i$, where $\lambda_i$
($\lambda_o$) and $\omega_i$  ($\omega_o$) are respectively the mean longitude  and
longitude of pericentre of the inner (outer) planet. Once the
resonance is established, the libration amplitude of $\psi$ slightly
increases with time until $t \sim 4\times 10^4$ binary orbits, and
then remains almost unchanged, suggesting that the planets are stably
locked into the resonance. From this time onward, the system is
close to an equilibrium state with both planets having constant
semimajor axes and eccentricities. At the end of the simulation, the
ratio of semimajor axes is $a_i/a_0 \sim 0.85$ and the ratio of
eccentricities is $e_i/e_o \sim 0.6$. Such a
configuration can be achieved because the torques exerted by the disc on
each planet act in an opposite way and can eventually counterbalance each other. From
the time 
the resonance is established  the negative
torques exerted by the disc  on the outer planet  make the two planets
migrate inward together. However, as both planets migrate, the innermost
planet experiences stronger positive corotation torques which
 tend to push the pair of planets outward. The bottom right panel of Fig. \ref{run1}  shows the evolution of the torques exerted on each planet as well as the effective torques acting on the whole system. We see that as the evolution proceeds, the  torques exerted on the inner planet are able to 
exactly counterbalance the ones exerted on the outer
body, which leads consequently to a zero net torque acting on the system. This  happens from 
$t\sim 2.1\times 10^4$, thereby stopping the joined migration of the planets.

\subsection{Models with $q\ge 1$}\label{qge1}
\begin{figure}
   \centering
   \includegraphics[width=\columnwidth]{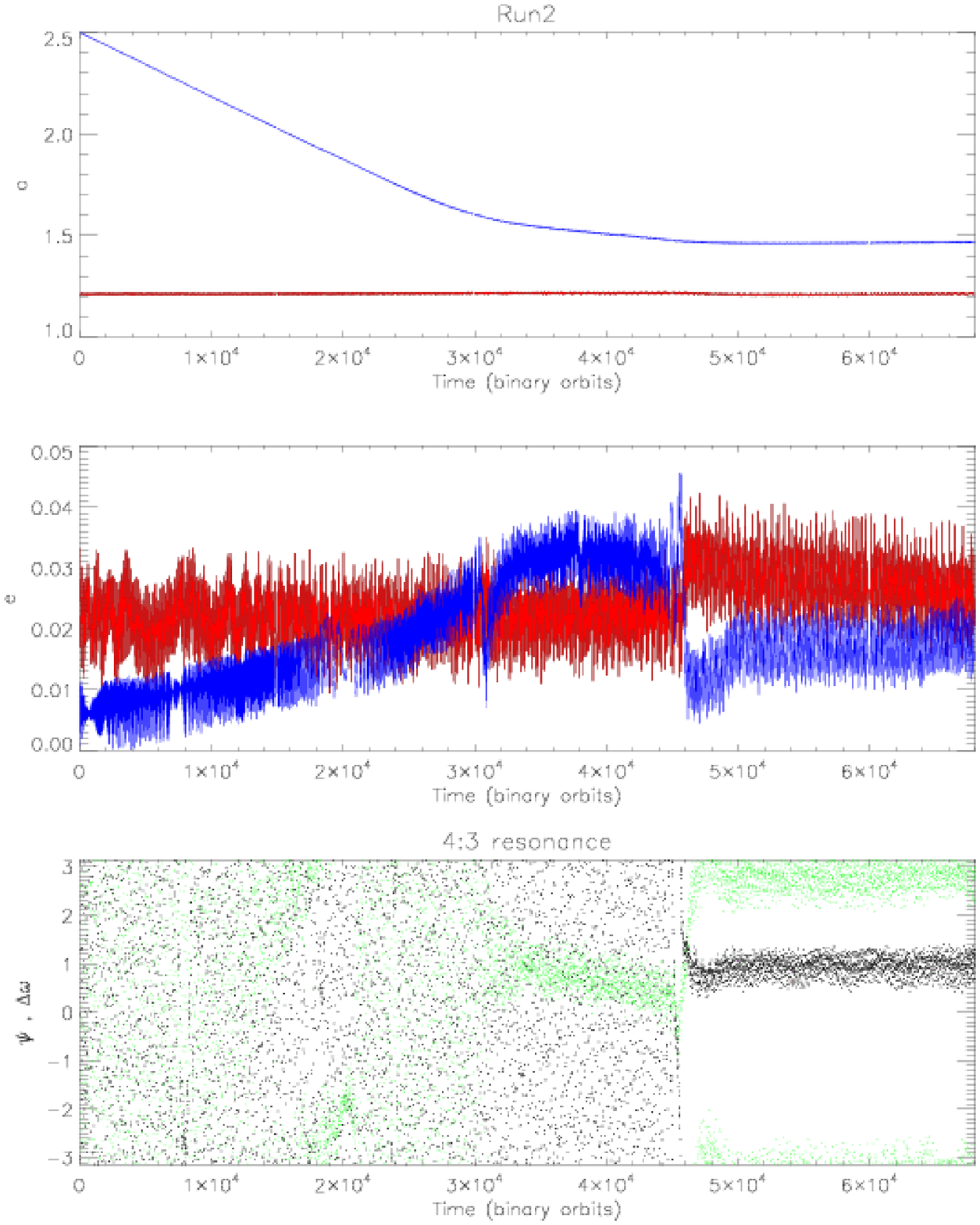}
      \caption{This figure shows the evolution of the semimajor
              axes and eccentricities for Run2 in which planets have masses of
              $m_i=10$ $M_\oplus$ (red line) and $m_o=5$ $M_\oplus$
              (blue line). The bottom panel displays the evolution of
              both the resonant angle
              $\psi=4\lambda_o-3\lambda_i-3\omega_i$ (black) associated with
              the 4:3 resonance and the apsidal angle
              $\Delta\omega=\omega_i-\omega_o$ (green).}
         \label{run2}
   \end{figure}

\begin{figure}
   \centering
   \includegraphics[width=\columnwidth]{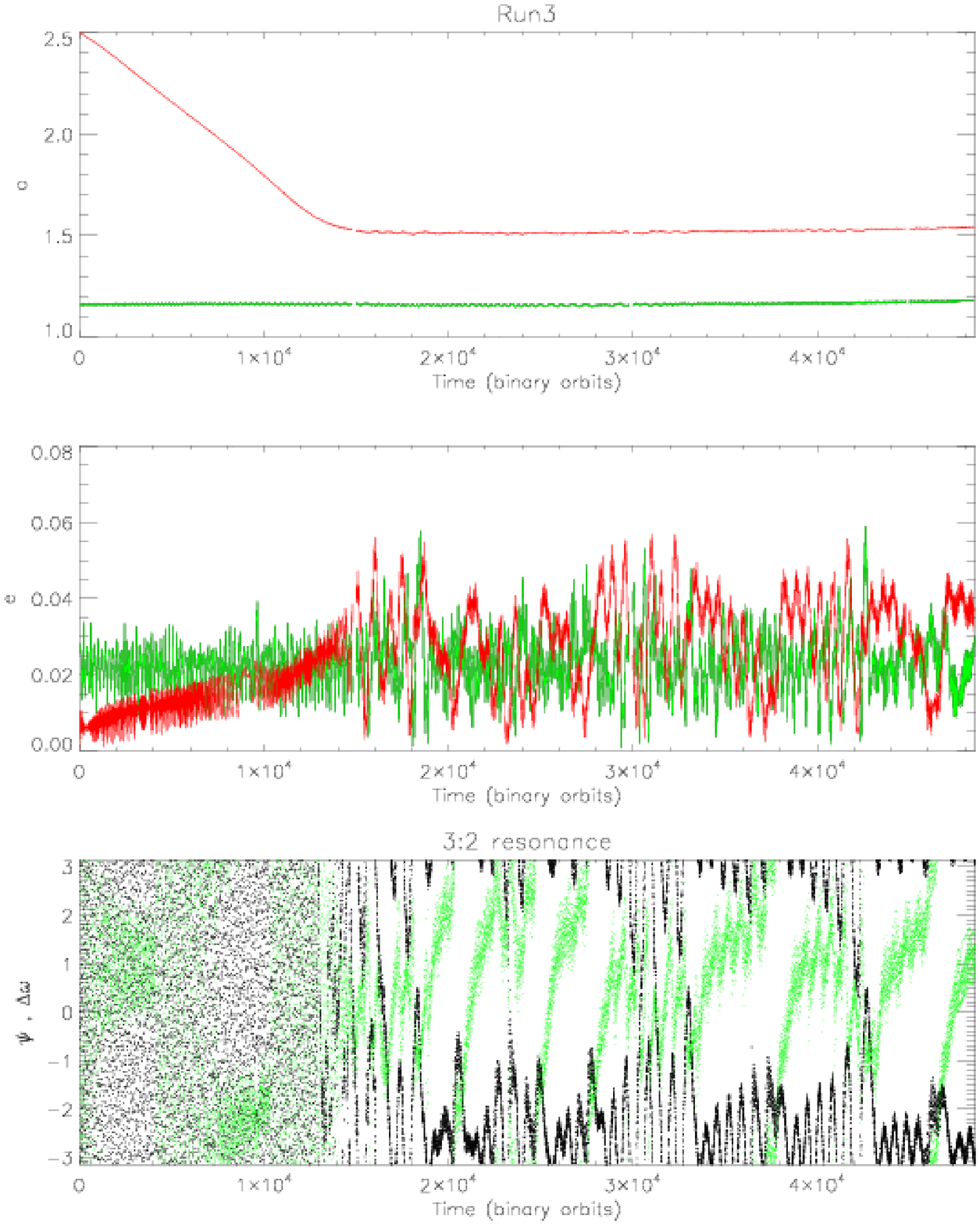}
      \caption{This figure shows the evolution of the semimajor
              axes and eccentricities for Run3 in which planets have masses of
              $m_i=20$  $M_\oplus$ (green line) and $m_o=10$
              $M_\oplus$ (red line). The bottom panel displays the
              evolution of both the resonant angle
              $\psi=3\lambda_o-2\lambda_i-2\omega_i$ (black) associated with
              the 3:2 resonance and the apsidal angle
              $\Delta\omega=\omega_i-\omega_o$ (green).}
         \label{run3}
   \end{figure}

\begin{figure}
   \centering
   \includegraphics[width=\columnwidth]{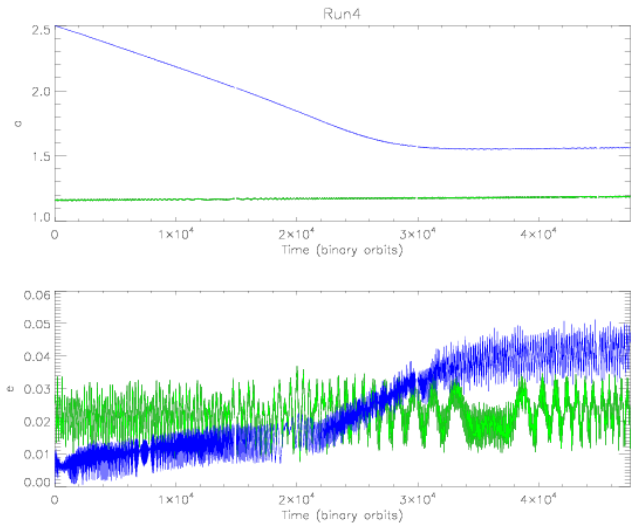}
      \caption{This figure shows the evolution of the semimajor
              axes and eccentricities for Run4 in which planets have masses of
              $m_i=20$  $M_\oplus$ (green line) and $m_o=5$ $M_\oplus$ (blue line).}
         \label{run4}
   \end{figure}

\begin{figure}
   \centering
   \includegraphics[width=\columnwidth]{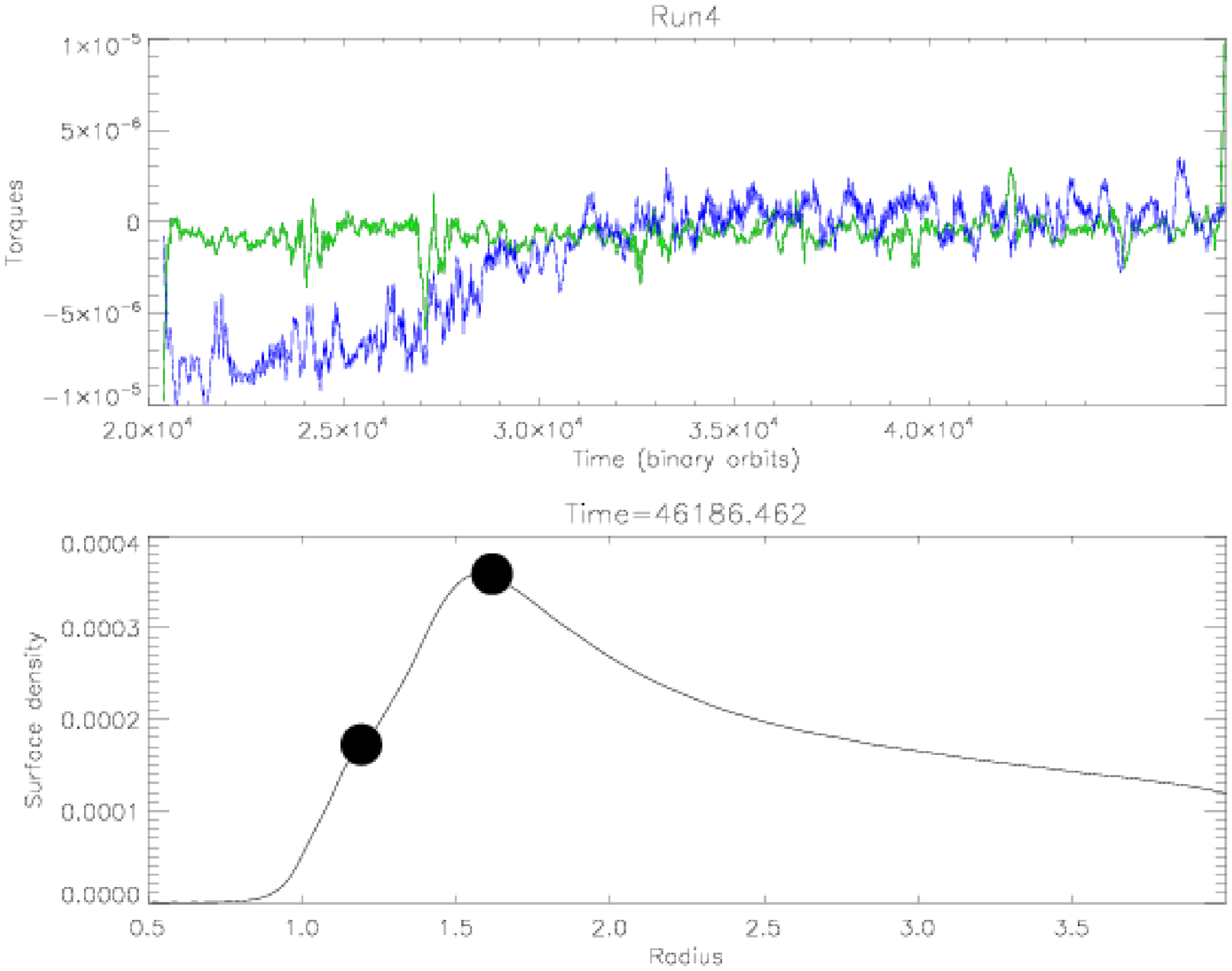}
     \caption{The upper panel shows the evolution of the torques
     exerted by the disc on the planets for Run4. The green line corresponds to the torques exerted  on the $20\;\mearth$ planet and the blue line corresponds to the torques
  exerted   on the $5\;\mearth$ planet. The lower panel displays
     the azimutal average of the surface density at the time shown
     above the plot as well as the position of the planets at the same
     time.}
         \label{run42}
   \end{figure}

Stable resonant locking was also found in some of the calculations with
$q \ge 1$. Fig. \ref{run2} shows the results for Run2 in which
planets have masses of $m_i=10$ and $m_o=5$ $\mearth$. With respect to
Run1,  the  outer planet migrates more slowly  since its mass is smaller.
However, the mode of evolution found in Run2 is very similar to the
one obtained in Run1, leading ultimately to a stable configuration with the two
planets trapped in the 4:3 resonance from  $t\sim 5\times 10^4$. At earlier times,  the evolution of $e_o$
   shows some peaks at $t\sim 1.9\times 10^4$ and  $t \sim 3\times
  10^4$ 
  which coincide with the planets being temporary captured in the 2:1 and 3:2
  resonances. At the end of the
 simulation, $e_o$ is still slightly increasing whereas $e_i$ is
 slightly decreasing, which indicates that the equilibrium
 configuration is not fully established. Nonetheless, comparing Figs.
 \ref{run1} and \ref{run2}, we can see that the libration amplitude of
 the resonant angle is much smaller in Run2 than in Run1,
 suggesting that the 4:3 resonance is more stable in this case.\\
In models with $m_i=20\;\mearth$,  the simulations  resulted in different modes of
evolution, depending on the mass of the outer planet
$m_o$. Fig. \ref{run3} and Fig. \ref{run4} display the results of
calculations with $m_o=10\;\mearth$ and $m_o=5\;\mearth$
respectively.  Comparing 
 these two figures, we can see that the final state
 of the system is quite similar in both cases, with planets evolving on fixed orbits with $a_i\sim 1.2$ and $a_o\sim 1.6$. In the run with $m_o=10\;\mearth$,
 a 3:2 resonance forms at $t\sim 1.4 \times 10^4$. In the simulation with  $m_o=5\;\mearth$ however, there is no evidence that the planets are in mean motion resonance, even though the system is close to the  3:2 commensurability. In this case, examination of the torques exerted by the disc (see upper panel of Fig. \ref{run42}) reveals that the migration of the system stalls because the torques acting on both planets cancel. Here, such an effect arises because the mass of the inner planet is high enough to
  make the onset of non-linear effects possible. These can
  significantly alter the surface density profile and widen the size of the inner cavity. Indeed,  we find that the edge of the inner cavity is located
  at $r\sim 1.2$ in simulations with  $m_i=10\;\mearth$ and $m_i=5\;\mearth$ whereas the bottom panel of
  Fig. \ref{run42} shows that it is located at $r\sim 1.5$ in Run4. 
Consequently, the evolution of the system in Run4 is such that the 
migration of the outermost planet is halted at the edge of the cavity 
formed by the binary plus inner planet system, therefore preventing 
capture in the 3:2 resonance (or in resonances of higher degree such as 4:3).   

\subsection{Models with $q < 1$}

\begin{figure}
   \centering
   \includegraphics[width=\columnwidth]{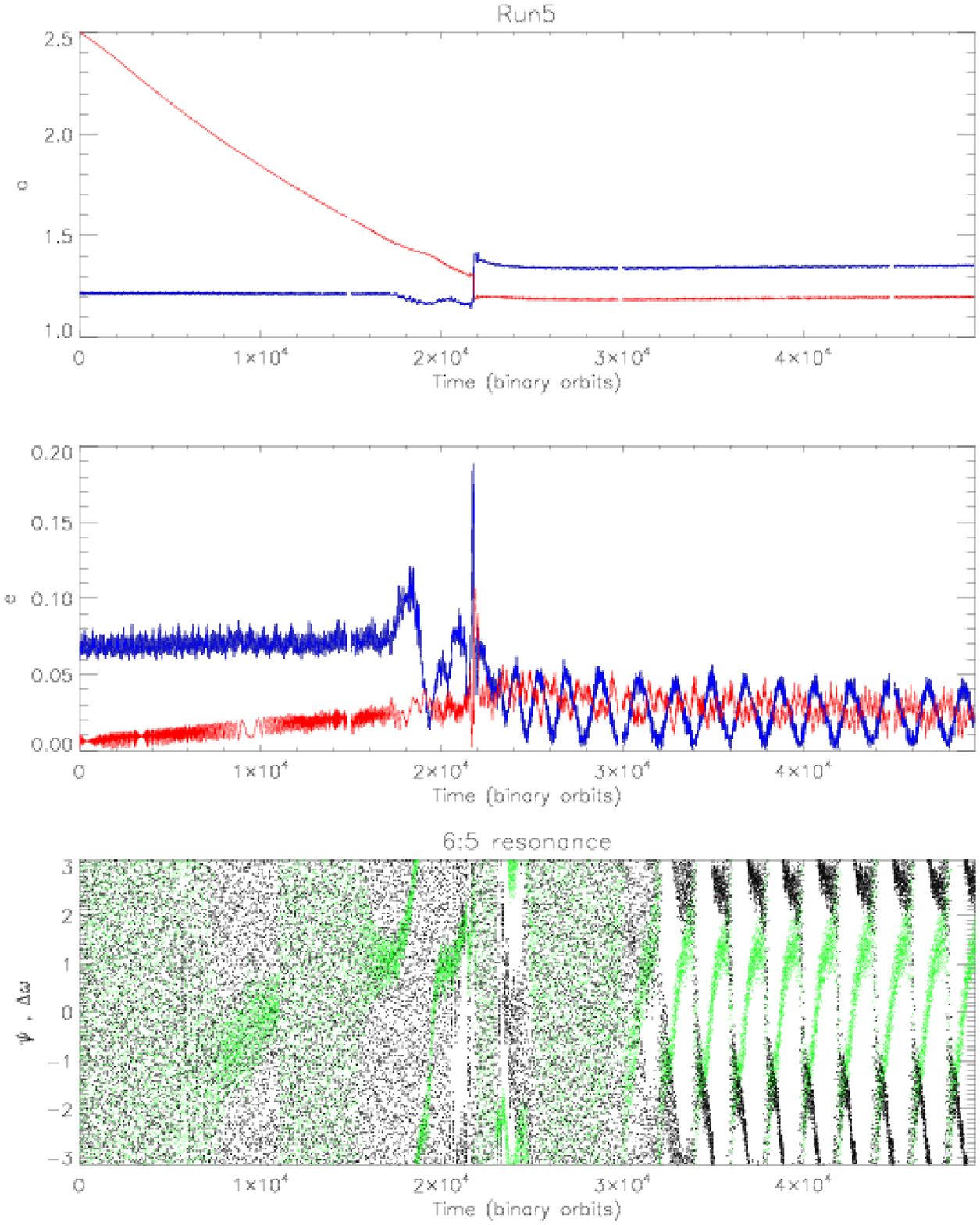}
     \caption{This figure shows the evolution of the semimajor
              axes and eccentricities for Run5 in which planets have masses of
              $m_i=5$  $M_\oplus$ (blue line) and $m_o=10$ $M_\oplus$
              (red line). The bottom panel displays the evolution of
              both the resonant angle
              $\psi=6\lambda_o-5\lambda_i-5\omega_i$ (black) associated with
              the 6:5 resonance and the apsidal angle
              $\Delta\omega=\omega_i-\omega_o$ (green).}
         \label{run5}
   \end{figure}

\begin{figure}
   \centering
   \includegraphics[width=\columnwidth]{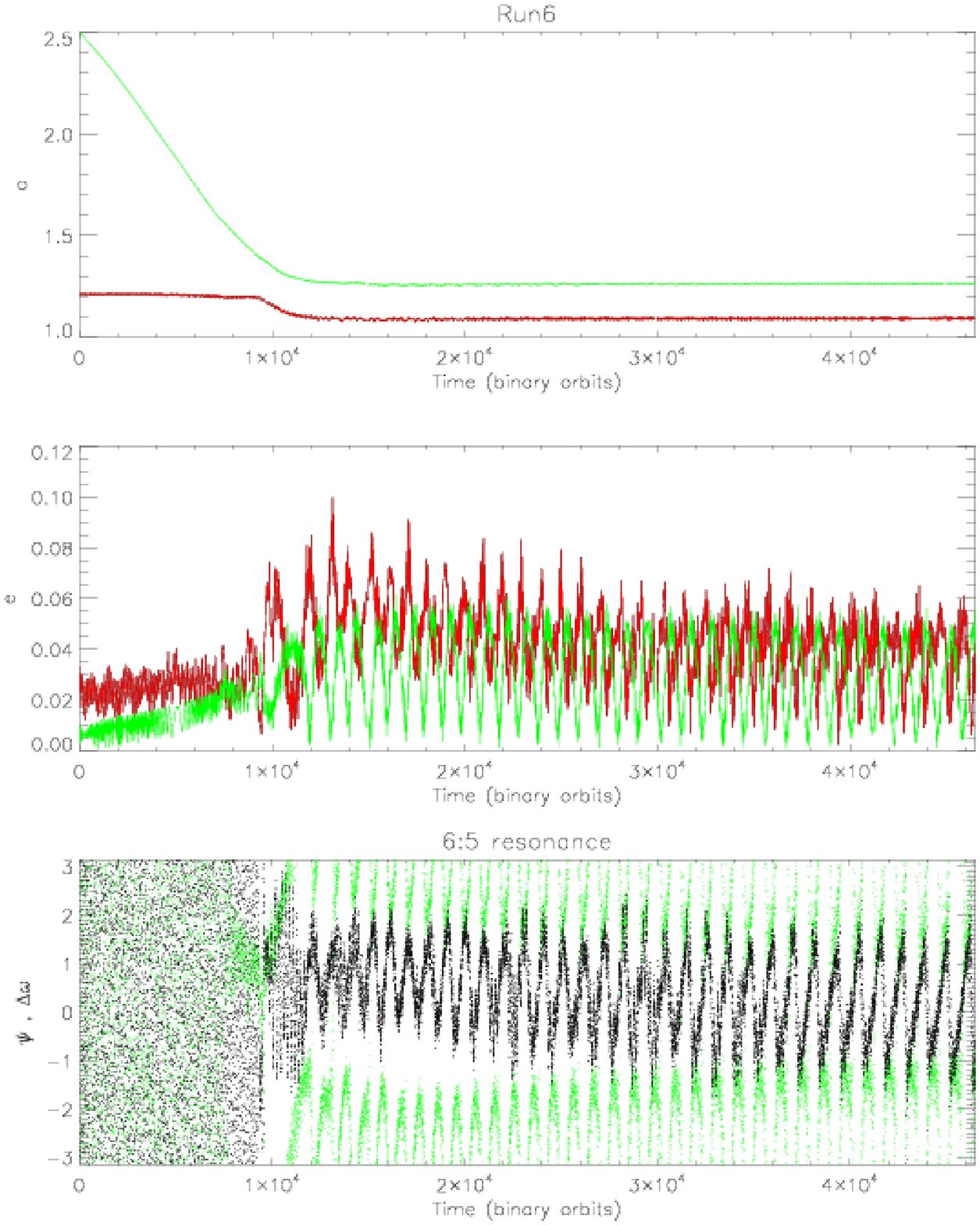}
      \caption{This figure shows the evolution of the semimajor
              axes and eccentricities for Run6 in which planets have masses of
              $m_i=10$  $M_\oplus$ (red line) and $m_o=20$ $M_\oplus$
              (green line). The bottom panel displays the evolution of
              both the resonant angle
              $\psi=6\lambda_o-5\lambda_i-5\omega_i$ (black) associated with
              the 6:5 resonance and the apsidal angle
              $\Delta\omega=\omega_i-\omega_o$ (green).}
         \label{run6}
   \end{figure}

\begin{figure}
   \centering
   \includegraphics[width=\columnwidth]{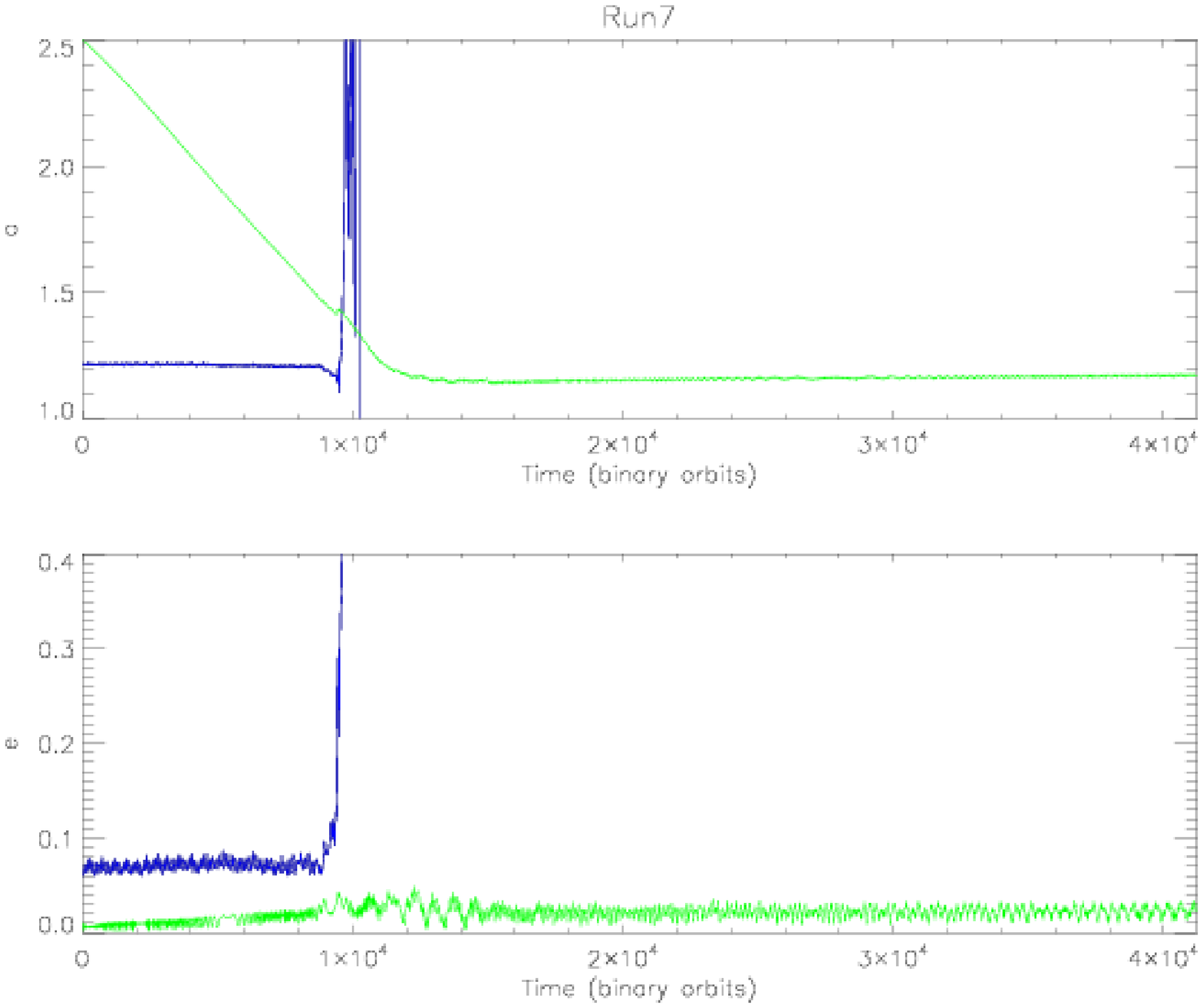}
      \caption{This figure shows the evolution of the semimajor
              axes and eccentricities for Run6 in which planets have masses of
              $m_i=5$ $M_\oplus$ (blue line) and $m_o=20$ $M_\oplus$ (green line).}
         \label{run7}
   \end{figure}

For simulations with $q < 1$, the evolution of the system differed
significantly from that just described in all but one case
(Run6). Fig. \ref{run5} shows the results of a simulation (Run5) with $q=0.5$
in which planets have masses of $m_i=5\;\mearth$ and $m_o=10\;\mearth$. Here, the outer planet passes
through the 4:3 resonance at $t\sim 1.8\times 10^4$  and
then slips into the 6:5 resonance with the inner planet at $t \sim 2\times
10^4$. The passage through these resonances is clearly
accompanied by an increase of the inner planet eccentricity $e_i$. At
$t\sim 2.1\times 10^4$, the inner planet undergoes a close encounter with the outer one as a result of resonant trapping. This subsequently leads to the scattering of the inner
planet further out in the disc while the outer one is  pushed inward
by virtue of conservation of angular momentum. Interestingly, such an orbital exchange leads to a configuration of the system similar to that of models  with $q\ge
1$. In good agreement with what we described in Section \ref{qge1}, we find that the final
state of the system is indeed an equilibrium configuration with the two
planets locked into the 6:5 resonance. At the end of the simulation, the
5  $\mearth$ and 10 $\mearth$ planets are respectively located at $a_i\sim
1.35$ and $a_o\sim 1.2$. \\
Although $q$ has the same value in Run6 as it is in Run5, we find a
different mode of evolution as shown in Fig. \ref{run6}.
In Run6 the planets have masses 
$m_i=10\;\mearth$ and $m_o=20\;\mearth$.
Relative to Run5 the disc induced eccentricity damping acting on the 
innermost planet is stronger, thereby preventing eccentricities from reaching
large values, 
and consequently preventing planets from undergoing close
encounters. Fig. \ref{run6} shows that the mode of evolution of the 
system in Run6 is
almost similar to that of models with $q\ge 1$, since the planets 
are in resonance (the 6:5 resonance in this case) and do not migrate. 
At the end of the evolution, the innermost  and outermost planets are
respectively located at $a_i\sim 1.1$  and $a_o\sim
1.3$  and the ratio of eccentricities is $e_i/e_o\sim
1.3$.\\
Fig. \ref{run7} shows the results of a calculation (Run7) with
$m_i=5\;\mearth$ and $m_o=20\;\mearth$, corresponding to a model with $q=0.25$.
Here, the 20 $\mearth$ planet enters the 4:3 resonance
with the 5 $\mearth$ body at $t\sim 9\times 10^3$, which
drives the eccentricity of the latter upward. This occurs until the
inner planet has a close encounter with the binary, resulting 
in the planet being completely ejected from the system. At later times, the
evolution is close to that described in Paper I, with
the $20\;\mearth$ planet migrating until it is trapped at the edge of
the cavity formed by the binary.

\section{Evolution of five-planet systems embedded in circumbinary discs}

We now  turn to the question of how a swarm of planets embedded in 
a circumbinary disc evolves. To address this issue, we have considered 
different models in which five planets having masses 
of 5, 7.5, 10, 12.5 and 15 $\mearth$ interact with each other. 
Three simulations have been performed in which we varied the initial 
configuration of the system. In one run (hereafter Model1), the initial mass 
distribution of planets decreases as one moves out in the disc,
whereas it is chosen to be random in Model2. In Model3 the initial 
mass distribution increases as a function of increasing orbital radius.

\subsection{Model1}

\begin{figure}
   \centering
   \includegraphics[width=\columnwidth]{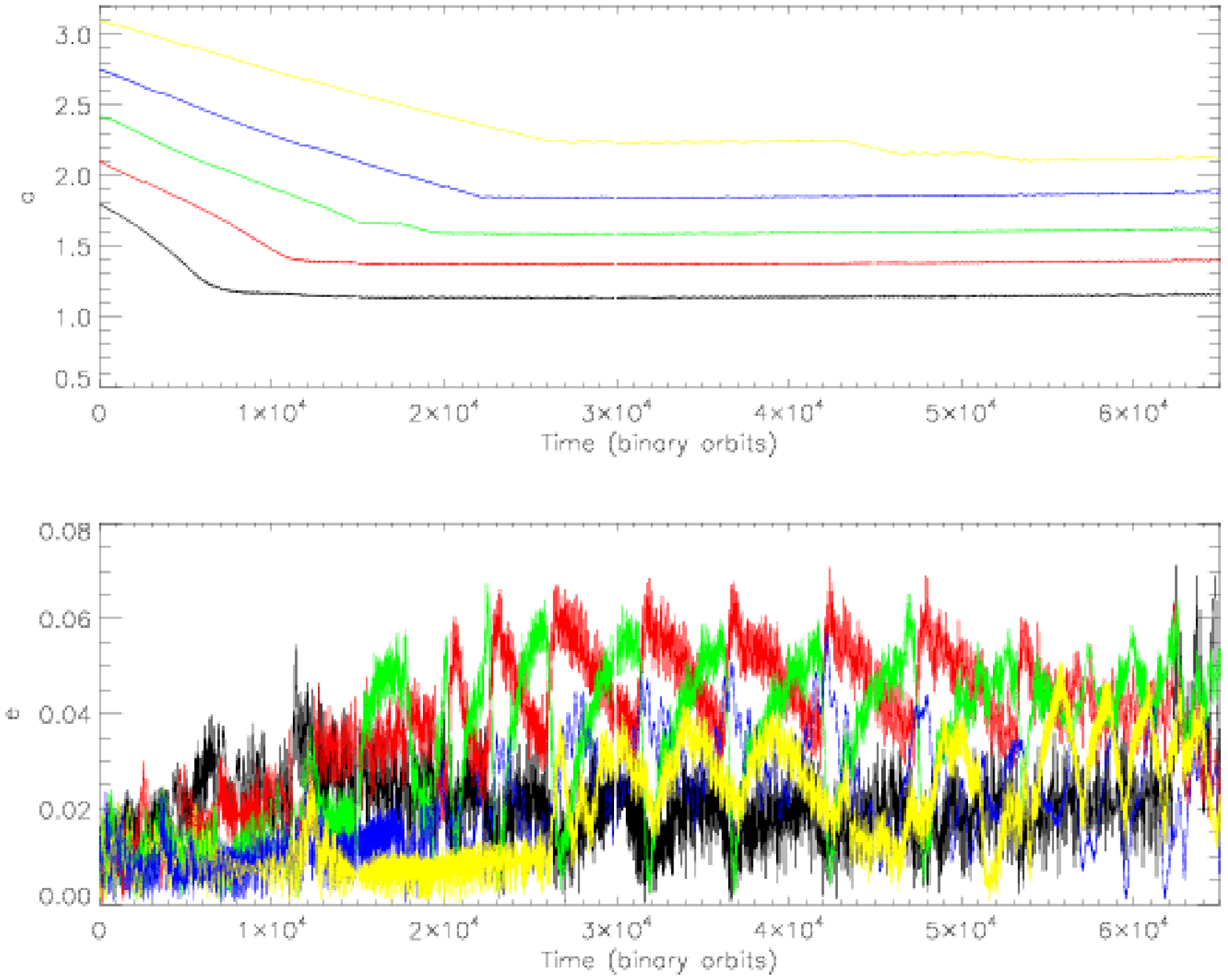}
      \caption{This figure shows the evolution of the semimajor
              axes and eccentricities of planets for Model1. Moving from the innermost planet to the outermost one, planets have masses of 15 (black line), 12.5 (red line), 10 (green line), 7.5 (blue line) and 5 $\mearth$ (yellow line). }
         \label{model1}
   \end{figure}

In all the simulations we have performed, the innermost planet initially 
evolves on a circular orbit with $a_1=1.8$. For this model
the initial planetary mass distribution decreases outward, 
so that the innermost 
planet has mass of $m_1=15\;\mearth$. As noted already, the planets
are spaced in orbital radius
assuming a mutual separation of $5\;R_{mH}$. Thus the other 
planets with masses of $m_p=$ 12.5, 10, 7.5, 5 $\mearth$ are 
located initially at $a_p=$ 2.1, 2.4, 2.8, 3.1 respectively.\\
Fig. \ref{model1} shows the evolution of the semimajor axes and eccentricities 
of planets for this model. At the beginning of the simulation, 
all the planets migrate inward as a consequence of type I migration, 
with a migration rate decreasing as one moves from the innermost planet to 
the outermost one. At $t\sim 7\times 10^3$, the innermost and most rapidly 
migrating $15\;\mearth$ planet reaches the edge of the inner  cavity
located at $r \sim 1.2$. As expected, this body remains trapped at 
this location until the inwardly migrating $12.5\;\mearth$ planet 
approaches and they enter the 4:3 resonance 
at $t\sim 1.1\times 10^4$. From this time the evolution of these 
two planets is similar to that of pairs of planets with $q\ge 1$ described 
in Section \ref{qge1}: the planets reach a quasi equilibrium state 
such that they evolve on non migrating orbits with  
$a_1\sim 1.2$ and $a_2\sim 1.4$, 
and with eccentricities remaining almost constant. This  lasts until the 
10 $\mearth$ planet enters the 4:3 resonance with the second body at 
$t\sim 1.5\times 10^4$. Again, the third planet tends to push the innermost 
planets inward, but the corotation torques exerted on the 15 $\mearth$ 
planet are able to counterbalance this effect and the migration of this 
three-planet system is stalled. A similar process occurs each time 
a migrating planet is resonantly captured by the bodies located inside 
its orbit. 
Over time we find that some planets slip from one resonance to another,
but the system remains globally stable during these episodes.
For example, the fifth planet with mass $m_5=5\;\mearth$ slips from the 4:3 
resonance with the $7.5\;\mearth$ planet to the 5:4 resonance at 
$t\sim 4.5\times 10^4$ and finally enters the 6:5 resonance at 
$t\sim 5.2\times10^4$. The final outcome of the simulation is a system 
forming a series of resonances between adjacent bodies with each of them 
evolving on a non migrating orbit. 

\begin{figure*}
   \centering
   \includegraphics[width=\columnwidth]{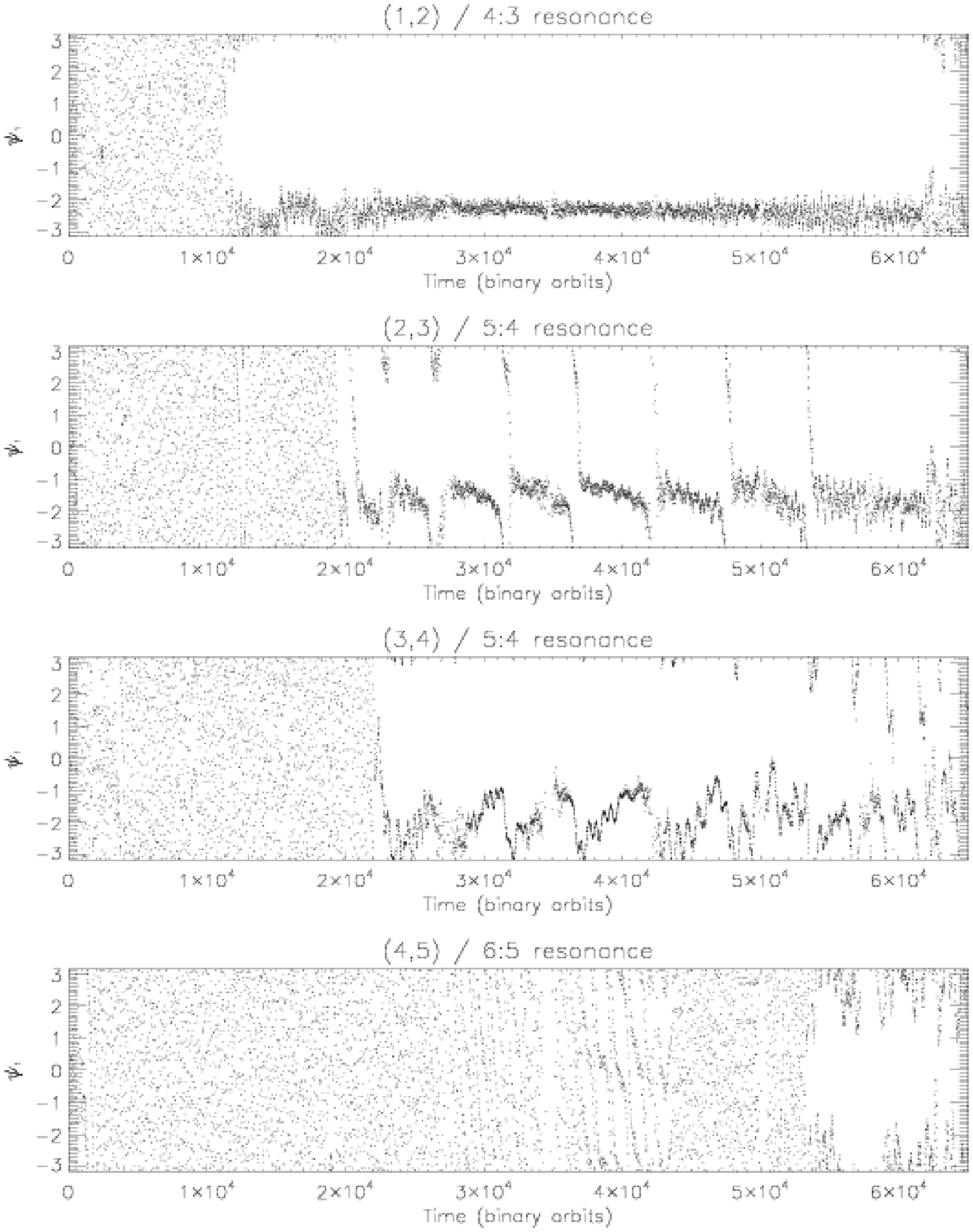}
   \includegraphics[width=\columnwidth]{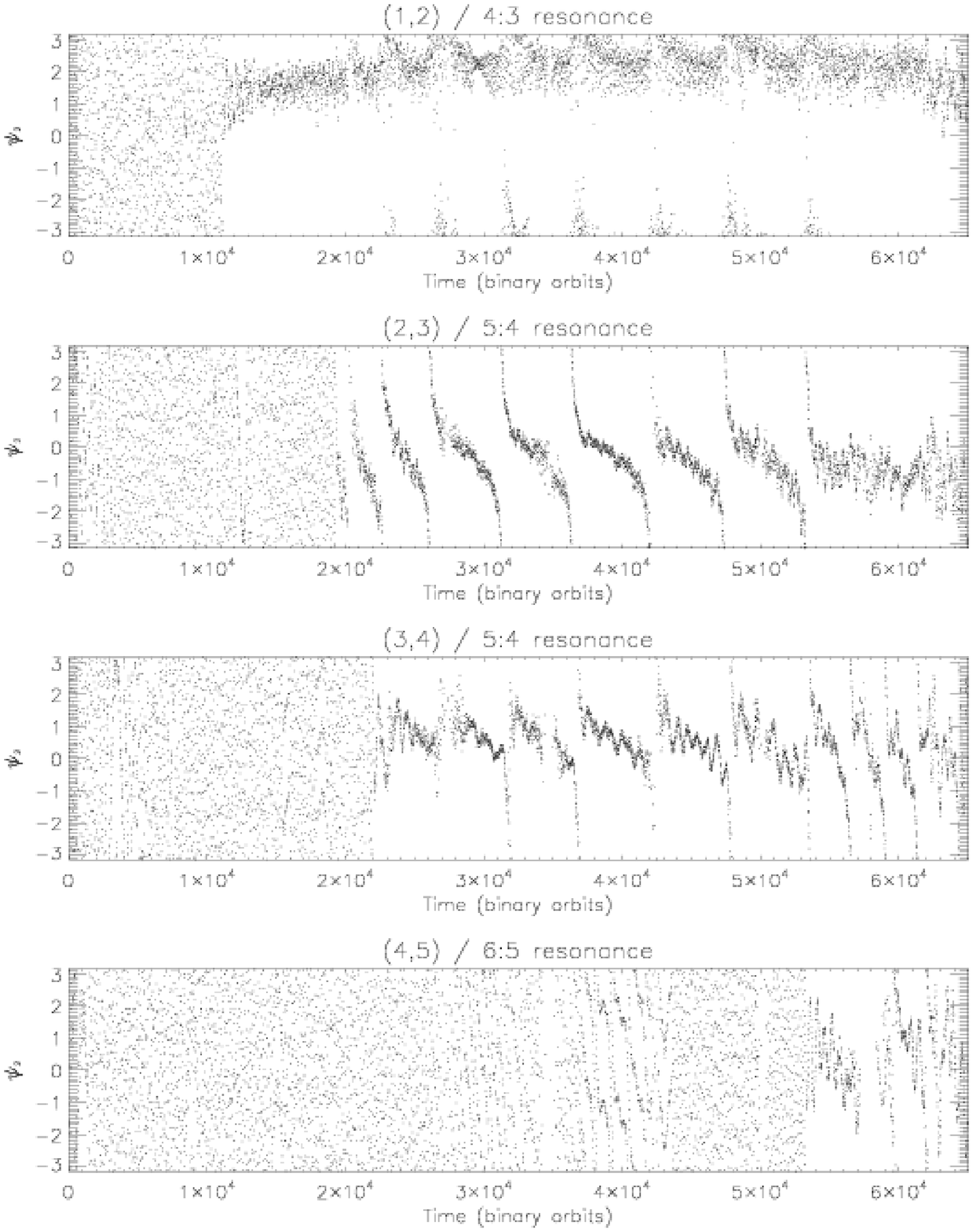}
      \caption{This figure shows the resonances which are established between adjacent bodies at the end of the simulation corresponding to Model1. Planets are labelled from 1 to 5, with 1 being the innermost planet and 5 being the outermost one.}
         \label{model1angles}
   \end{figure*}

Fig.\ref{model1angles} displays the resonant 
angles  $\psi_1=(p+1)\lambda_o-p\lambda_i-\omega_i$ and
  $\psi_2=(p+1)\lambda_o-p\lambda_i-\omega_o$ corresponding to the $(p+1):p$ commensurabilities that form between each pair of
adjacent bodies. We see that all commensurabilities that form are first order 
resonances, in agreement with results obtained by  
Cresswell \& Nelson (2006). This simulation suggests that in a 
circumbinary disc,  corotation torques exerted at the edge of the inner 
cavity provide an efficient mechanism against type I migration for a
swarm of planets, and that resonant capture prevents close encounters,
scattering and collisions when the initial planetary mass
distribution decreases as a function of orbital radius.

\subsection{Model2}
\begin{figure}
   \centering
   \includegraphics[width=\columnwidth]{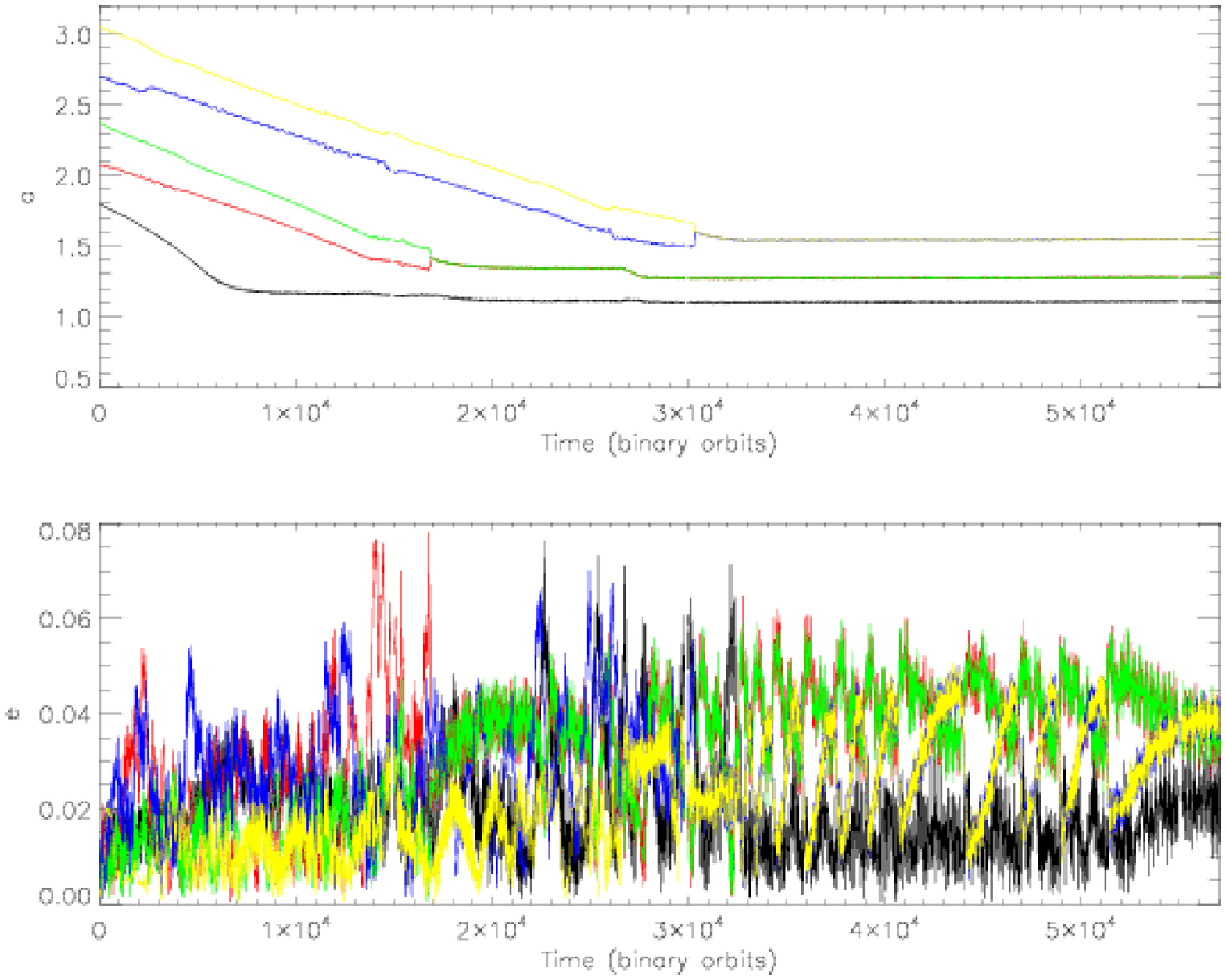}
      \caption{This figure shows the evolution of the semimajor
              axes and eccentricities of planets for Model2. Moving from the innermost planet to the outermost one, planets have masses of 15 (black line), 7.5 (red line), 12.5 (green line), 5 (blue line) and 10 $\mearth$ (yellow line).}
         \label{model2}
   \end{figure}

\begin{figure*}
   \centering
   \includegraphics[width=8cm]{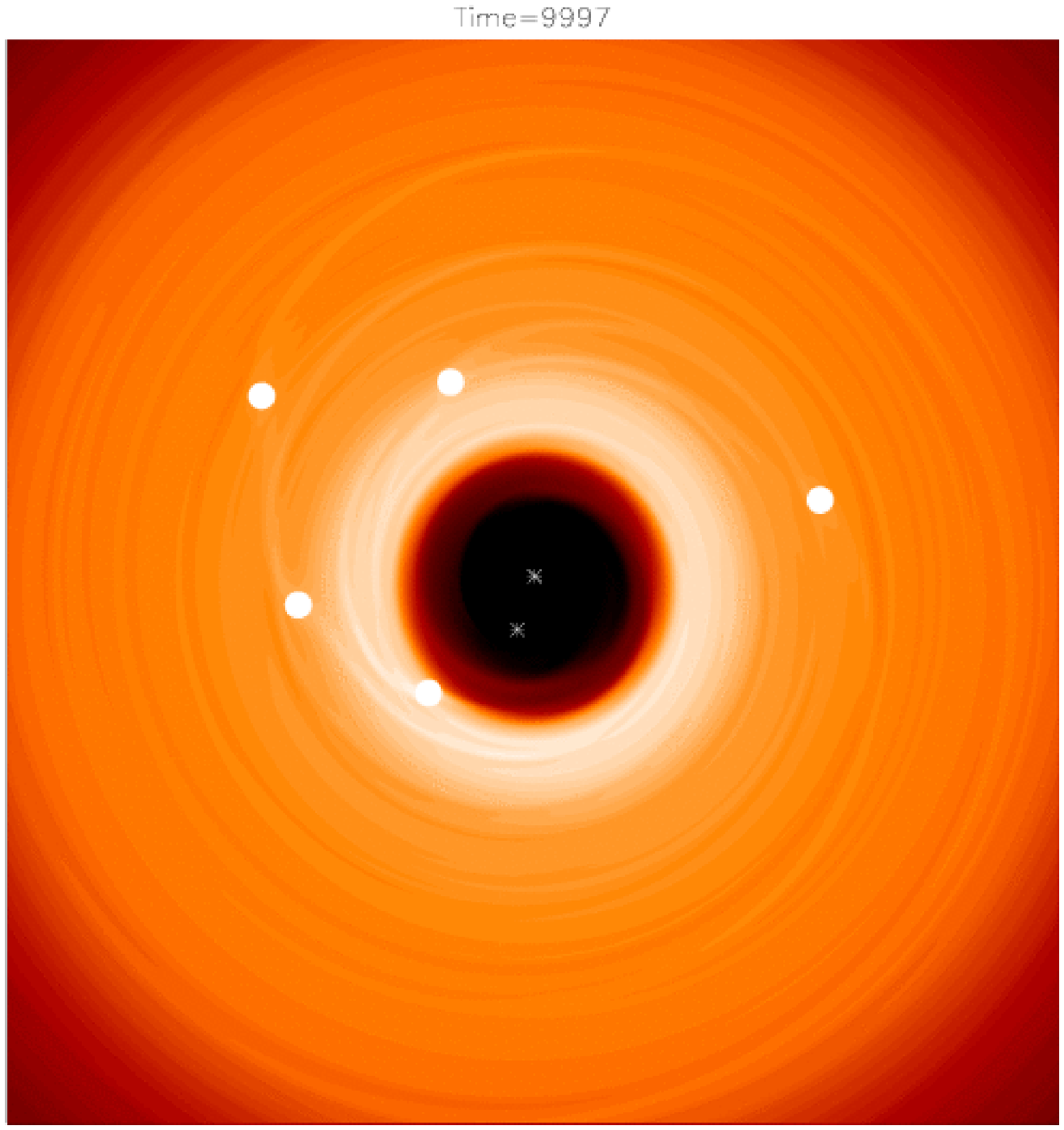}
    \includegraphics[width=8cm]{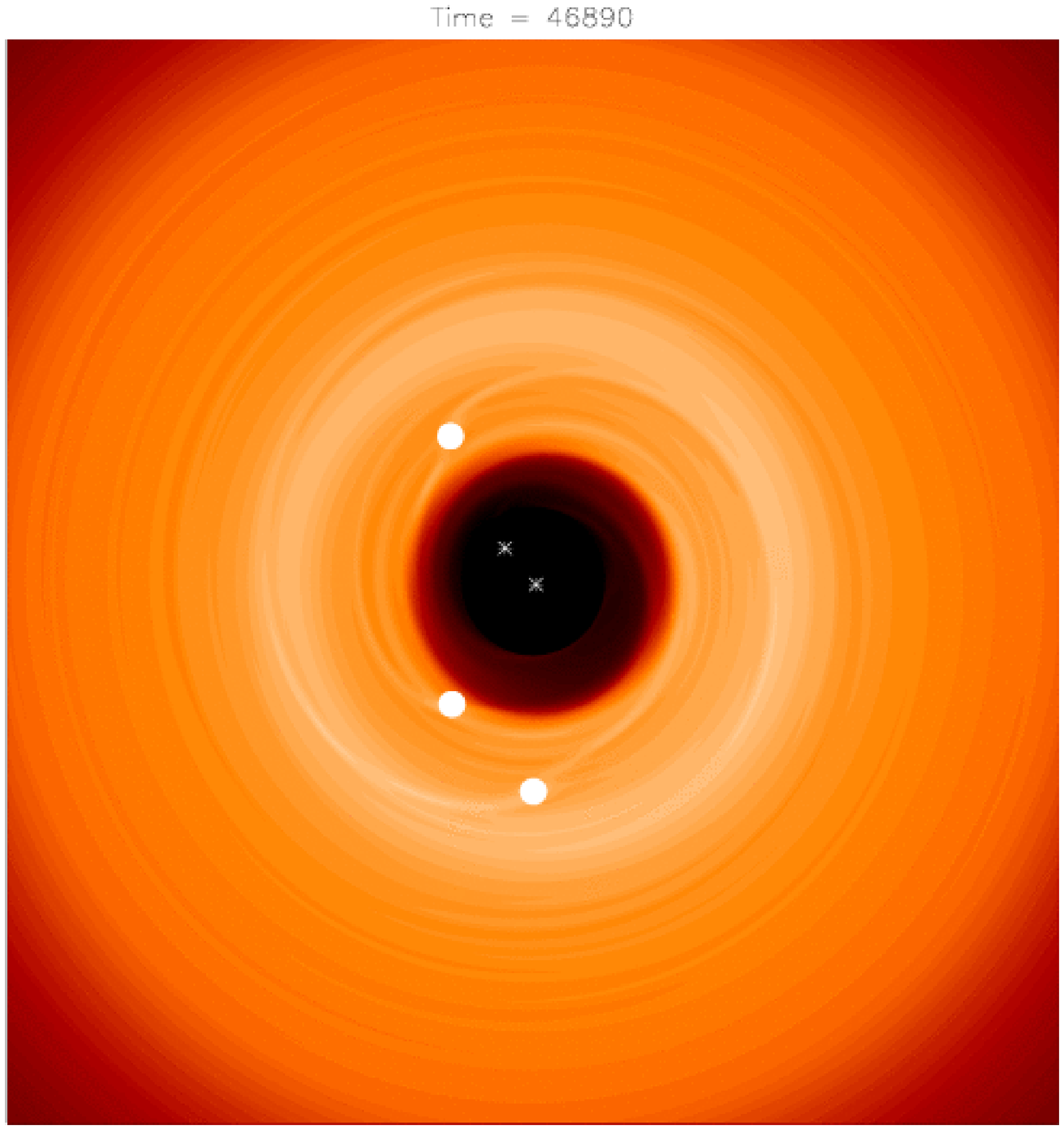}
      \caption{This figure shows, for Model2, snapshots of the disc surface density at times shown above the plots. In this figure, planets are represented by white circles.}
         \label{model22d}
   \end{figure*}

In this model, the initial configuration of the system is such that
moving from the innermost planet to the outermost one, planets have masses 
of $m_p=$ 15, 7.5, 12.5, 5 and 10 $\mearth$ respectively. \\
A snapshost of the disc surface density at the beginning of the simulation 
is presented in the left panel of Fig. \ref{model22d} and the evolution of the 
semimajor axes and eccentricities of planets for this model is 
illustrated in Fig. \ref{model2}. Once again, the innermost 15 $\mearth$ 
planet rapidly drifts toward the edge of the inner cavity where it remains 
trapped. In comparison with Model1, adjacent bodies have significantly 
different masses which leads to a stronger diffential migration between them.
This causes the 12.5 $\mearth$  and 10 $\mearth$ planets to rapidly catch up 
with the 7.5 $\mearth$ and 5 $\mearth$ planets respectively, leading 
to the formation of resonances. Here, 
the 12.5 $\mearth$ planet first enters the 6:5 resonance with the 
7.5 $\mearth$ body and then slips into the 7:6 resonance. 
Fig. \ref{model2} shows that the excitation of eccentricities due to this 
resonant interaction, and the influence of surrounding planets,
 leads ultimately to a collision between these two 
bodies at $t\sim1.7\times10^4$. This merger 
forms a 20 $\mearth $ planet, which then migrates inward
until it becomes locked stably into the 
5:4 resonance with the 15 $\mearth$ planet. At $t\sim 3\times 10^4$ a new 
collision occurs between the 5 and 10 $\mearth$ planets.
Earlier, these two planets passed through a 
sequence of different resonances and were in the 7:6 resonance just before this
collision occured. Again, this newly formed 15 $\mearth$ planet migrates 
until it catches with the 20 $\mearth$ body. 

\begin{figure*}
   \centering
   \includegraphics[width=\columnwidth]{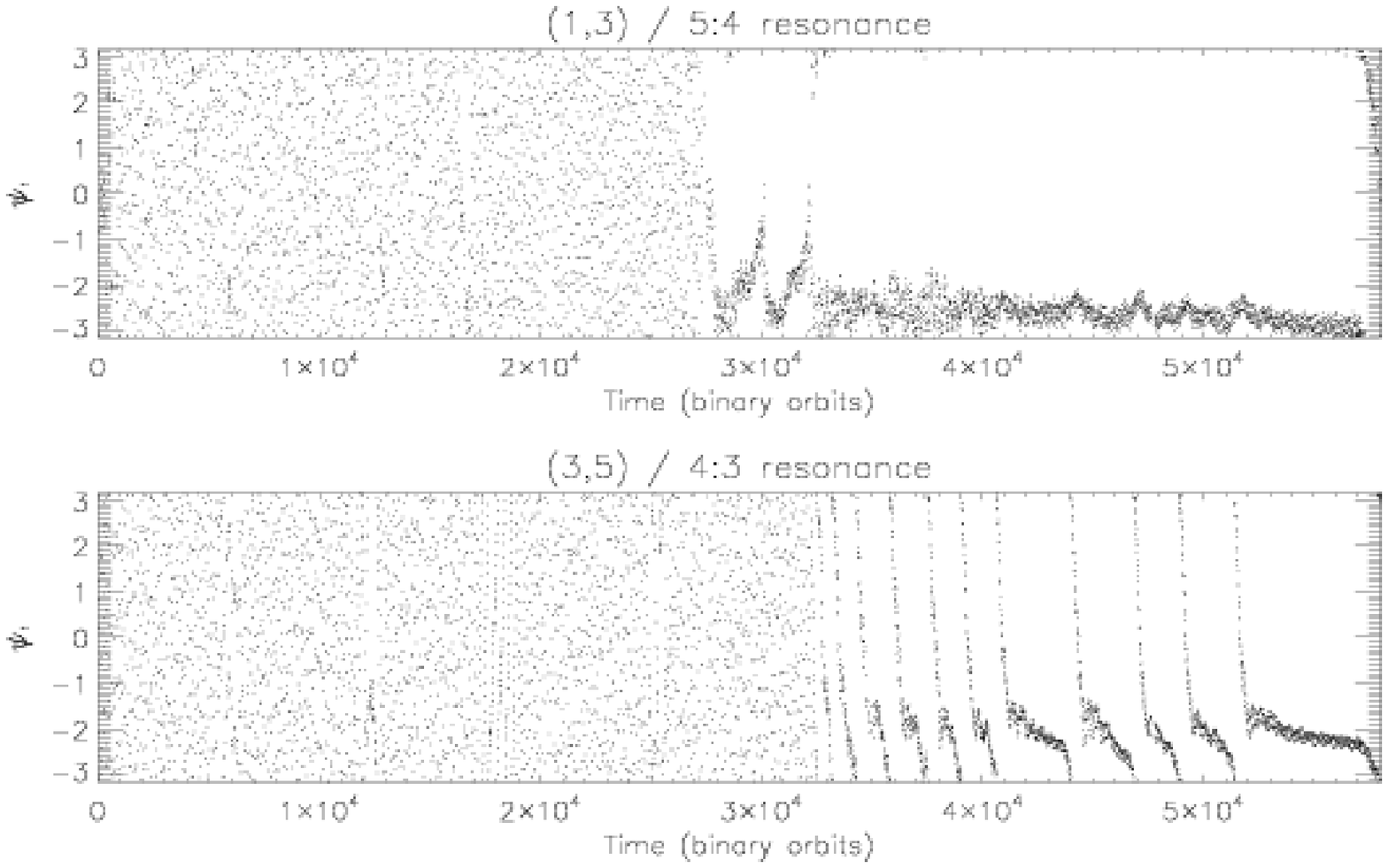}
   \includegraphics[width=\columnwidth]{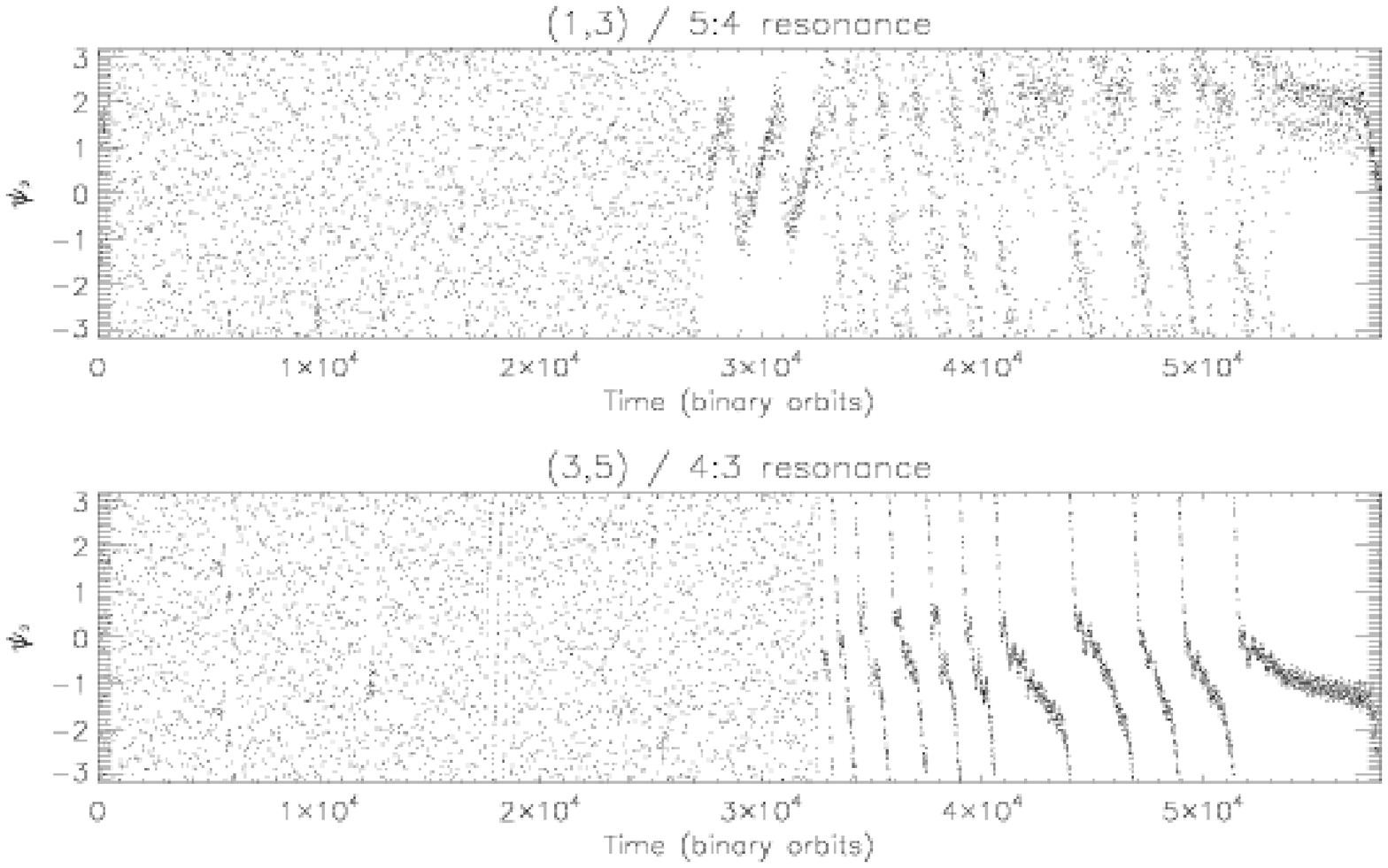}
      \caption{This figure shows the resonances which are established between adjacent bodies at the end of the simulation corresponding to Model2. Planets are labelled from 1 to 5, with 1 being the innermost planet and 5 being the outermost one.}
         \label{model2angles}
   \end{figure*}

The right panel of 
Fig. \ref{model22d} shows a snapshot of the disc surface density at the 
end of the simulation. We see that the final state of the system consists 
of only three bodies evolving in the disc. Moving from the innermost planet 
to the outermost one, these have masses of 15, 20 and 15 $\mearth$. Once 
again, the corotation torques exerted at the cavity edge prevent inward 
migration, leaving each surviving body in resonance with its
neighbours. Fig. \ref{model2angles} displays the resonant angles corresponding
to the commensurabilities that form between each pair. These are 
the 5:4 resonance for the first pair and the 4:3 resonance for the second one.
This is in good agreement with the simulations performed by 
Cresswell \& Nelson (2006) who studied the evolution of a swarm of 
planets embedded in a protoplanetary disc, and who found that the 
4:3, 5:4, 6:5 and 7:6 resonances are most favoured. 

\subsection{Model3}

\begin{figure}
   \centering
   \includegraphics[width=\columnwidth]{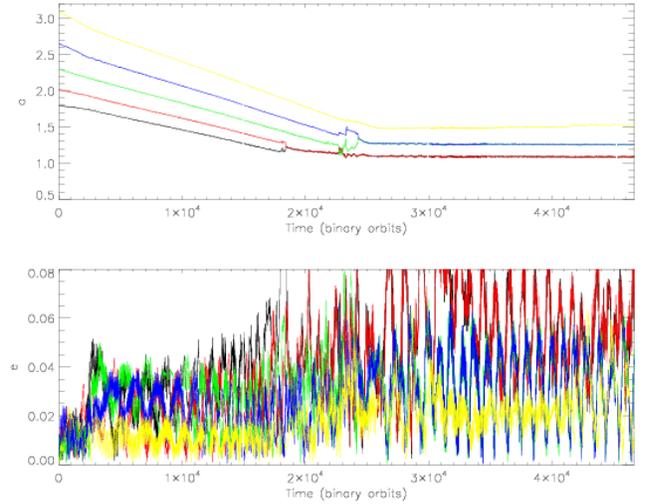}
      \caption{This figure shows the evolution of the semimajor
              axes and eccentricities of planets for Model3. Moving from the innermost planet to the outermost one, planets have masses of 5 (black line), 7.5 (red line), 10 (green line), 12.5 (blue line) and 15 $\mearth$ (yellow line).}
         \label{model3}
   \end{figure}

In this last model, we consider planets with masses 
$m_p=$ 5, 7.5, 10, 12.5 and 15 $\mearth$ located initially at 
$a_p=$ 1.8, 2, 2.3, 2.7 and 3.1 respectively.\\
The evolution of the semimajor axes and eccentricities of planets for 
Model3 are shown in Fig. \ref{model3}. At the beginning of the simulation,
each planet migrates inward, but the initial configuration of this model 
is such that the orbits of two adjacent bodies rapidly converge, leading 
to the formation of resonances. At $t\sim3\times 10^3$, the 
system has already evolved to a state where each body is in resonance with 
its closest neighbours, which occurs here before the innermost planet
reaches the inner cavity. Moving from the inner planet to the outer one, 
the commensurabilities that form  are respectively the 
7:6, 6:5, 6:5 and 5:4 resonances. The resonant 
interaction between the 5 and 7.5 $\mearth$ planets increases
their eccentricities up to $e_p\sim 0.08$. Eccentricity growth within the swarm
leads to crossing orbits, and subsequent interactions cause the breaking
of resonances. This leads to a collison between the 5 and 7.5 $\mearth$
planets at $t \sim 1.8 \times 10^{4}$.
At $t\sim 2.4\times 10^4$ a collision resulting in a 
22.5 $\mearth$ planet also occurs between the 10 and 12.5 $\mearth$ bodies 
which were locked into the 7:6 resonance prior to this event. 

\begin{figure}
   \centering
   \includegraphics[width=\columnwidth]{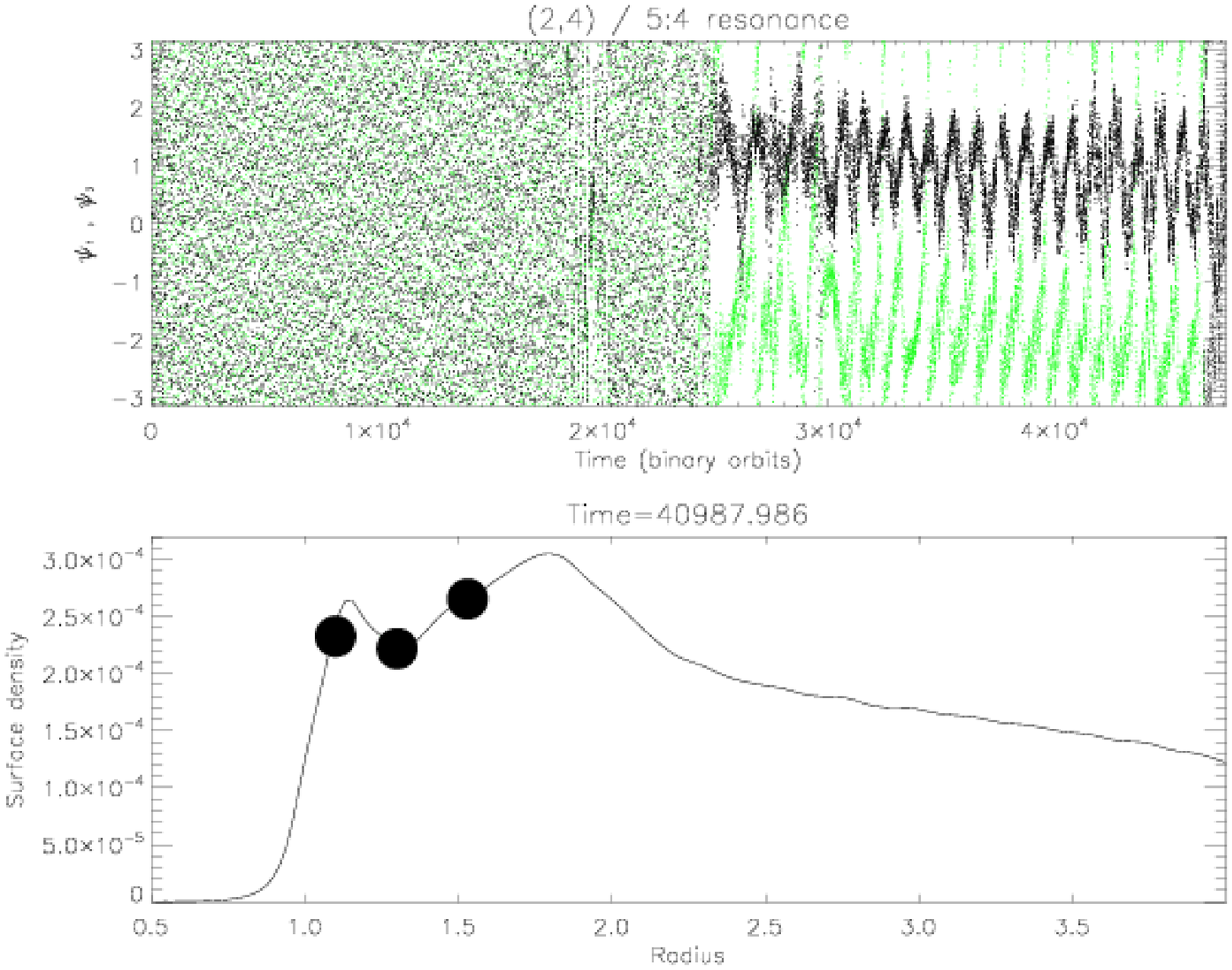}
      \caption{The upper panel shows, for Model3, the resonant angles
      $\psi_1$ (black) and $\psi_2$ (green) associated with the resonance that forms between the two innermost planets. The bottom panel displays the  disc surface density profile at the time shown above the plot. The semimajor axes of the planets at that time are represented by black circles.}
         \label{model3angles}
   \end{figure}

The final state of the system consists of three planets 
with  masses of $m_p=$ 12.5, 22.5 and 15 $\mearth$ respectively located at
$a_p=$ 1.1, 1.3 and 1.5. The upper panel of Fig. \ref{model3angles} shows 
that the 5:4 resonance is clearly established between the first pair of 
planets. At the end of the simulation, these two bodies evolve on 
non migrating
orbits, whereas the outermost body  migrates outward very slightly, 
indicating that the latter is not in resonance with its interior neighbour.  
The disc surface density profile at $t\sim 4\times 10^4$ is displayed in the 
bottom panel of Fig. \ref{model3angles}. It shows that the outermost planet 
is located in a region where the disc has a large positive surface density 
gradient, close to the outer 
edge of the partial gap formed by the 22.5 $\mearth$ body. 
This results in large 
positive corotation torques, leading to the observed outward 
migration of the outermost planet. This process is likely to operate until the 
planet reaches a fixed point located near the gap edge, where the total 
torque (corotation plus Lindblad) cancels (Masset et al. 2006).
This result is consistent with that found in Section \ref{qge1} where 
the inner and outer planet masses were 20 and 5 $\mearth$, respectively.

\subsection{Long term evolution}

We now turn to the question of the long-term evolution of the planetary
systems obtained in the five-planet simulations. 
Because the interaction with the gas
disc tends to damp eccentricities, it is necessary to
examine the dynamical stability of the planets after the disc
dispersal to establish long term stability. 
Each of the previous simulations was restarted at a
point corresponding to the end of the run, but with the gas
surface density decaying exponentially with an e--folding time 
$t_{dec}= 2\times 10^3$.
Once these systems had evolved for $\sim 10^4$ binary orbits, 
by which time the surface density in the discs had decreased
by a factor of $\sim 10^3$, we continued the simulations
with a pure N-body code, ignoring any residual effects of the
remaining gas.
For each of the five-planet models the results
of this procedure are presented in Fig. \ref{model1}, which displays
the time evolution of the orbital radii of the planets. 

In Model1, the
eccentricity growth resulting from the disc dispersal gives
rise, at the beginning of the simulation, to numerous scattering events 
that eventually lead to collisions. At time $t\sim 8\times 10^4$
 a system of two planets with masses of 27.5 and 22.5 $\mearth$ remains,
but these merge at $t\sim 1.1\times 10^5$. The final state of the
system is then a 50
$\mearth$ planet orbiting at $r\sim 1.5$. 

A similar outcome
is obtained in Model3 in which the long-term evolution resulted in a
15 $\mearth$ planet evolving in a high-eccentricity orbit with a
semi-major axis of $a_p\sim 2$. At earlier times, the increase in
eccentricities following disc dispersal 
led to a collision between the 12.5 and 22.5 $\mearth$
bodies, thereby forming a new 35 $\mearth$ planet. At $t\sim 2\times
10^5$, the latter is observed to undergo a close encounter with the
cental binary, leading to this body being completely ejected
from the system. \\
Interestingly, the three-planet system in Model2 appears to be
dynamically stable over long time scales, with the planets maintaining
their commensurabilities. This indicates that multiplanet resonant
systems could potentially be found in circumbinary discs, where the
existence of the resonance helps to maintain the stability of the system.

\begin{figure}
   \centering
   \includegraphics[width=\columnwidth]{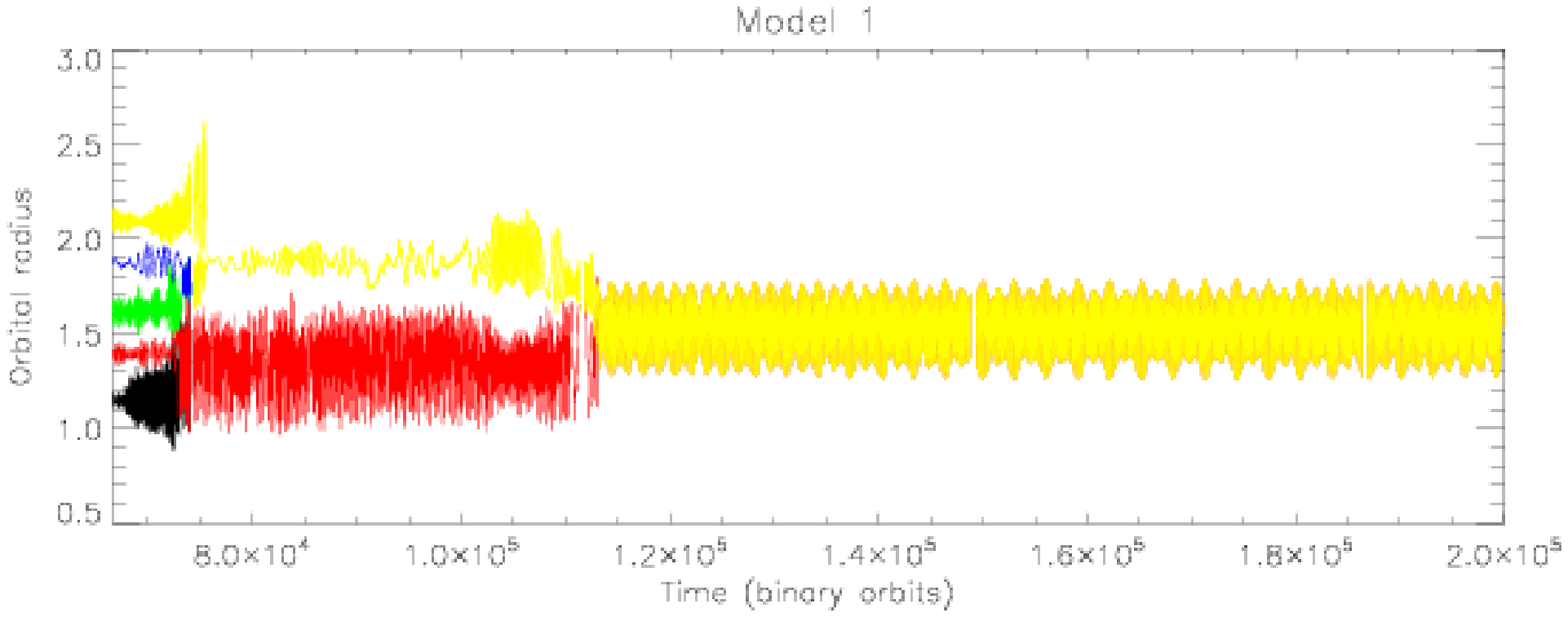}
   \includegraphics[width=\columnwidth]{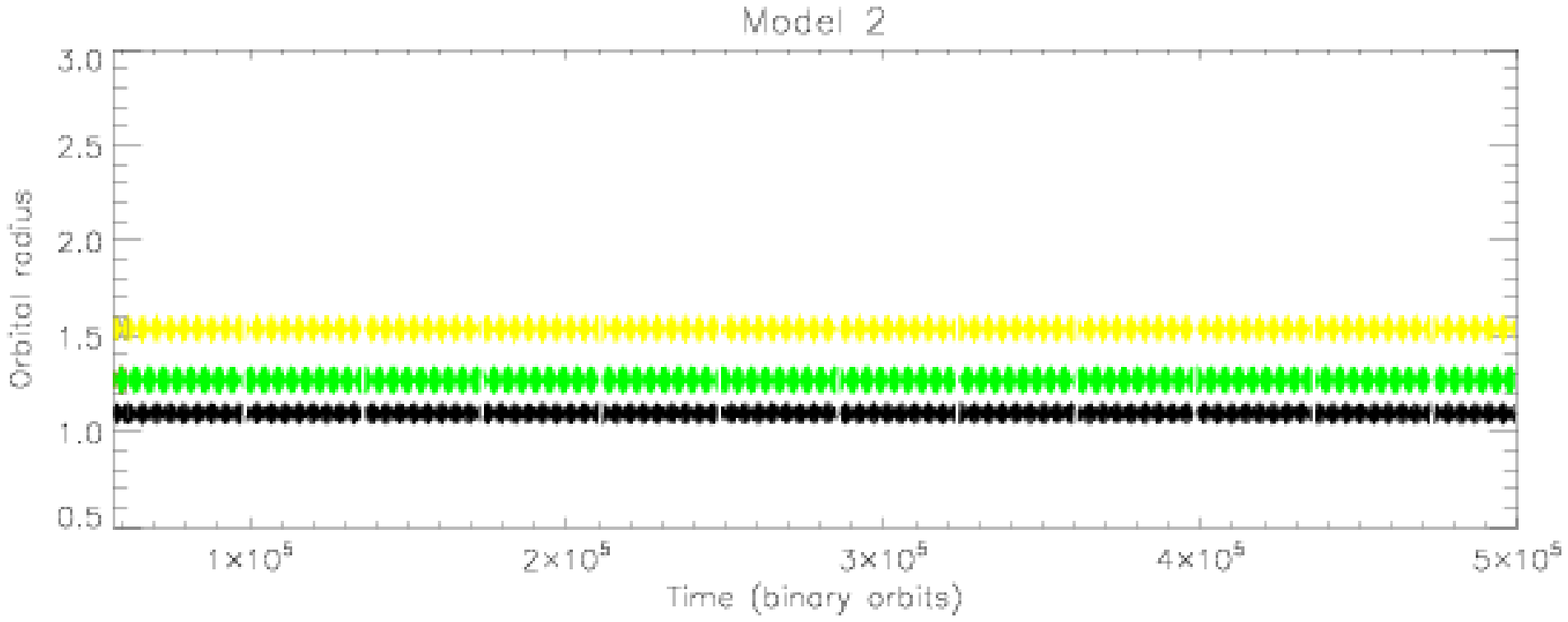}
   \includegraphics[width=\columnwidth]{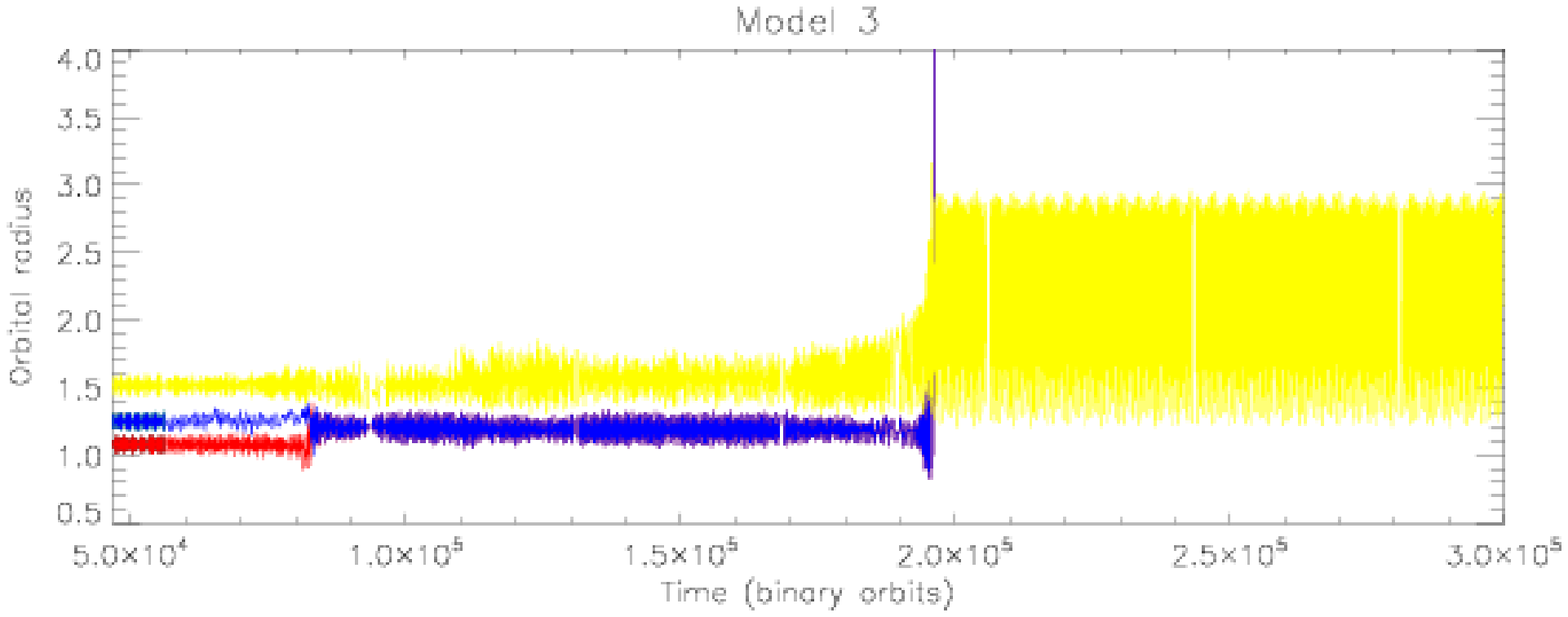}
      \caption{This figure shows time evolution of the orbital radii
      of the planets for Model1, Model2 and Model3 as a consequence of
      the disc dispersal. It corresponds to the results of hydro plus
      N-body simulations (see text for details). }
         \label{nbody}
   \end{figure}
\section{Summary and conclusion}

In this paper we have presented the results of hydrodynamic simulations 
aimed at studying the evolution of multiple planets embedded in a 
circumbinary disc. \\
We first focused on a system consisting of a pair of planets 
interacting with each other. We assumed that one body is trapped at 
the edge of the inner cavity formed by the binary, while the other migrates 
inward from outside the orbit of the innermost body. 
Our calculations show different outcomes, depending on the planet mass 
ratio $q=m_i/m_o$.\\
i) For models with $q=1$, the simulations indicate that the system evolves 
toward a quasi equilibrium state with both planets trapped in a mean 
motion resonance and evolving on non migrating orbits. This occurs 
because the positive corotation torques exerted on the innermost planet 
counterbalance the negative corotation torques exerted on the outermost one.\\
ii) For models with $q \ge 1$, most of the runs resulted in a similar mode of 
evolution. Occasionally however, the final fate of the sytem was such that 
the two bodies are in close vicinity to one another, have stopped migrating,
but are not in resonance. This arises when the mass of the innermost planet 
is high enough to open a partial gap in the disc, such that the migration 
of the outer planet is stopped at the edge of the large cavity formed 
by both the binary and the innermost planet.\\
iii) For most of the models with $q < 1$, the planets involved in the 
simulations underwent different dynamical processes such as scattering or 
orbital exchange. When orbital exchange occurs, the final state of the system is
a stable mean motion resonance with the more massive planet now being the 
innermost one. Scattering and orbital exchange occurs once a first order 
resonance is established between the planets. This drives up the 
eccentricity of the inner planet leading to a close encounter with 
either the outer planet or the binary. A close encounter with the
binary leads to ejection.

Having examined the two-planet problem, we then
focused on  more ``realistic'' systems composed of five
planets with masses  of 5, 7.5, 10, 12.5 and 15 $\mearth$
embedded in a circumbinary disc. We performed three simulations,
with the planets being placed in different initial orbital configurations.

In general terms, the evolution of such a system
proceeds as follows. Each planet migrates
inward until it is captured into resonance with an adjacent
body. This occurs either because the innermost planet has reached the
edge of the inner cavity, where it remains trapped, or because two
adjacent bodies migrate differentially. In some simulations, 
resonant interaction and the associated eccentricity growth
resulted in  close encounters and collisions between bodies,
resulting in the formation of a more massive planet.
At later times, once the more chaotic phase of evolution is over,
the system behaves more quiescently, and
generally reaches a stationary state with
each remaining body in resonance with its closest neighbours
on non migrating orbits due to planet trapping at the cavity edge.
In two of the three simulations, we found that disc dispersal
led to strong scattering and collisions between bodies, resulting in eventually
only one planet being left in the system.
In the remaining case, we found that three planets could exist
stably in  a three body mean motion resonance over very long times.

The results of our simulations indicate that resonant systems of
planets with masses in the $\sim$ few--Earth mass range may 
be common in close binary systems. Indeed, the inner edge of a 
circumbinary disc plays the role of a barrier, which raises the 
possibility that two bodies become resonantly trapped. This
resonance trapping, however, usually prevents the two planets from approaching
each other so that they are unable to merge to form a larger body.
When more than two bodies are introduced we found that the 
resulting scattering and close encounters could lead to planetary growth,
however, but we have been unable to run sufficient numbers 
of simulations to provide reliable statistical information about the range
of outcomes. 

A number of outstanding issues remain when considering planet formation
in circumbinary discs. The first is understanding where the planetary cores
can form due to planetesimal accretion. The formation of an eccentric
disc around a binary system, and the presence of the binary itself,
can increase the velocity dispersion to a value where accretion
does not occur except far out in the disc. 
The second is understanding the growth of cores
once they have formed, either through continued planetesimal
accretion, or through giant impacts. The migration of the core toward the
inner cavity may actually place it in a region where planetesimal
accretion is reduced due to increased velocity dispersion,
and at present there have been no simulations of giant impact
growth which include the gas disc. If a giant core can be formed, however,
then it is likely that a gas giant planet will result.
As it grows, a giant planet should leave the edge of the cavity and 
undergo type II migration (e.g. Lin \& Papaloizou 1993; Nelson \& al. 2000), 
until it becomes trapped into the 4:1 resonance with the binary. 
This appears to be the most likely fate of a giant planet migrating in a
circumbinary disc (Nelson 2003), although trapping is not necessarily
permanent and simulations indicate a finite probablity of the
planet being ejected from the system due to close encounters with the binary.
The evolution of the planet during
its growth from a  core into a gas giant, however,
has not been explored, and this will be the subject of a future paper.

\begin{acknowledgements}
The simulations performed in this paper were performed on
the QMUL High Performance Computing facility purchased
under the SRIF iniative.
\end{acknowledgements}

\end{document}